\documentclass{spie}

\usepackage{fixltx2e}
\usepackage[T1]{fontenc}
\usepackage[utf8]{inputenc}
\usepackage{bold-extra}
\usepackage{xspace}
\usepackage{color}
\usepackage{graphicx}
\usepackage{multirow}
\usepackage{tabularx}
\usepackage{array}
\usepackage{amsfonts}
\usepackage{amssymb}
\usepackage{xfrac}
\usepackage{tikz}

\newcolumntype{C}{>{\centering\arraybackslash}X}%

\usepackage{aas_macros}

\newcommand{\planck}{{\itshape Planck}\xspace}
\newcommand{\spider}{{\scshape Spider}\xspace}

\newcommand{\ukrts}{$\mu$K$\sqrt{s}$\xspace}
\newcommand{\icm}{cm$^{-1}$\xspace}

\title{Pre-flight integration and characterization of the {\bfseries\scshape\LARGE Spider} balloon-borne telescope}

\affilcounter{caltech}{Division of Physics, Mathematics and Astronomy, California Institute of Technology, Pasadena, CA, USA}
\affilcounter{cardiff}{School of Physics and Astronomy, Cardiff University, Cardiff, UK}
\affilcounter{cifar}{Canadian Institute for Advanced Research, Toronto, ON, Canada}
\affilcounter{cita}{Canadian Institute for Theoretical Astrophysics, University of Toronto, Toronto, ON, Canada}
\affilcounter{cwru}{Department of Physics, Case Western Reserve University, Cleveland, OH, USA}
\affilcounter{durban}{School of Mathematics, Statistics and Computer Science, University of KwaZulu-Natal, Durban, South Africa}
\affilcounter{ias}{Institut d'Astrophysique Spatiale, Orsay, France}
\affilcounter{imperial}{Theoretical Physics, Blackett Laboratory, Imperial College, London, UK}
\affilcounter{jpl}{Jet Propulsion Laboratory, Pasadena, CA, USA}
\affilcounter{kavli}{Kavli Institute for Cosmology, University of Cambridge, Cambridge, UK}
\affilcounter{kipac}{Kavli Institute for Particle Astrophysics and Cosmology, SLAC National Accelerator Laboratory, Menlo Park, CA, USA}
\affilcounter{nist}{National Institute of Standards and Technology, Boulder, CO, USA}
\affilcounter{northwestern}{CIERA, Northwestern University, Evanston, IL, US}
\affilcounter{princeton}{Department of Physics, Princeton University, Princeton, NJ, USA}
\affilcounter{princetonastro}{Department of Astrophysical Sciences, Princeton, NJ, USA}
\affilcounter{stanford}{Department of Physics, Stanford University, Stanford, CA, USA}
\affilcounter{toronto}{Department of Physics, University of Toronto, Toronto, ON, Canada}
\affilcounter{torontoastro}{Department of Astronomy and Astrophysics, University of Toronto, Toronto, ON, Canada}
\affilcounter{ubc}{Department of Physics and Astronomy, University of British Columbia, Vancouver, BC, Canada}

\authorcounter[\noaffil$\dagger$,princeton]{A.\,S.\ Rahlin}
\authorinfo{$^\dagger$Corresponding author. Email: arahlin@princeton.edu. \\
Copyright 2014 Society of Photo-Optical Instrumentation Engineers. One print or electronic copy may be made for personal use only. Systematic reproduction and distribution, duplication of any material in this paper for a fee or for commercial purposes, or modification of the content of the paper are prohibited.}
\authorcounter[cardiff]{P.\,A.\,R.\ Ade}
\authorcounter[ubc]{M.\ Amiri}
\authorcounter[toronto]{S.\,J.\ Benton}
\authorcounter[caltech,jpl]{J.\,J.\ Bock}
\authorcounter[cita,cifar]{J.\,R.\ Bond}
\authorcounter[cwru]{S.\,A.\ Bryan}
\authorcounter[durban]{H.\,C.\ Chiang}
\authorcounter[imperial]{C.\,R.\ Contaldi}
\authorcounter[caltech,jpl]{B.\,P.\ Crill}
\authorcounter[caltech,jpl]{O.\ Dor\'e}
\authorcounter[toronto,cita]{M.\ Farhang}
\authorcounter[caltech]{J.\,P.\ Filippini}
\authorcounter[toronto,northwestern]{L.\,M.\ Fissel}
\authorcounter[princeton]{A.\,A.\ Fraisse}
\authorcounter[princeton]{A.\,E.\ Gambrel}
\authorcounter[torontoastro]{N.\,N.\ Gandilo}
\authorcounter[caltech]{S.\ Golwala}
\authorcounter[princeton]{J.\,E.\ Gudmundsson}
\authorcounter[ubc,cifar]{M.\ Halpern}
\authorcounter[princetonastro,ubc]{M.\,F.\ Hasselfield}
\authorcounter[nist]{G.\ Hilton}
\authorcounter[jpl]{W.\,A.\ Holmes}
\authorcounter[caltech]{V.\,V.\ Hristov}
\authorcounter[stanford,kipac,nist]{K.\,D.\ Irwin}
\authorcounter[princeton]{W.\,C.\ Jones}
\authorcounter[princeton]{Z.\,D.\ Kermish}
\authorcounter[stanford,kipac]{C.\,L.\ Kuo}
\authorcounter[kavli]{C.\,J.\ MacTavish}
\authorcounter[caltech]{P.\,V.\ Mason}
\authorcounter[jpl]{K.\ Megerian}
\authorcounter[caltech]{L.\ Moncelsi}
\authorcounter[caltech]{T.\,A.\ Morford}
\authorcounter[cwru]{J.\,M.\ Nagy}
\authorcounter[torontoastro,toronto,cifar]{C.\,B.\ Netterfield}
\authorcounter[caltech,jpl]{R.\ O'Brient}
\authorcounter[nist]{C.\ Reintsema}
\authorcounter[cwru]{J.\,E.\ Ruhl}
\authorcounter[jpl]{M.\,C.\ Runyan}
\authorcounter[torontoastro]{J.\,A.\ Shariff}
\authorcounter[ias,torontoastro]{J.\,D.\ Soler}
\authorcounter[jpl]{A.\ Trangsrud}
\authorcounter[cardiff]{C.\ Tucker}
\authorcounter[caltech]{R.\,S.\ Tucker}
\authorcounter[jpl]{A.\,D.\ Turner}
\authorcounter[jpl]{A.\,C.\ Weber}
\authorcounter[ubc]{D.\,V.\ Wiebe}
\authorcounter[princeton]{E.\,Y.\ Young}
\begin{document}

\maketitle

\begin{abstract} We present the results of integration and characterization of
the \spider instrument after the 2013 pre-flight campaign. \spider is a
balloon-borne polarimeter designed to probe the primordial gravitational wave
signal in the degree-scale $B$-mode polarization of the cosmic microwave
background. With six independent telescopes housing over 2000 detectors in the
94 GHz and 150 GHz frequency bands, \spider will map 7.5\% of the sky with a
depth of 11 to 14 $\mu$K$\cdot$arcmin at each frequency, which is a factor of
$\sim$5 improvement over \planck. We discuss the integration of the pointing,
cryogenic, electronics, and power sub-systems, as well as pre-flight
characterization of the detectors and optical systems. \spider is well prepared
for a December 2014 flight from Antarctica, and is expected to be limited by
astrophysical foreground emission, and not instrumental sensitivity, over the
survey region.

\noindent\textbf{Keywords:} \spider, cosmic microwave background, polarization,
inflation, transition-edge sensor, scientific ballooning,
millimeter wave instrumentation, cosmology.
\end{abstract}

\section{Introduction}

The past twenty years have seen the field of cosmology expand at a rapid pace,
from the first hints of anisotropies in the Cosmic Microwave Background (CMB)
detected by COBE\cite{smoot1992}, to precision tests of the Standard
Cosmological Model by the latest generation of CMB
experiments\cite{bicep2_spect, pbear_bb, planck2013_parameters, wmap9params,
  act2013_parameters_all, actpol2014ee, sptsz_all} and complementary datasets,
such as late-time measurements of the scale of baryon-acoustic
oscillations\cite{boss2012_all}, and direct measurements of the Hubble
parameter\cite{riess2011}.

The CMB traces the distribution of matter in the photon-baryon fluid of the
early universe, and anisotropies in this distribution are the seeds of structure
formation that evolve into the galaxies and clusters that we observe today. The
same density perturbations that generate the temperature anisotropies also
generate a polarization pattern on the sky by means of Thomson scattering within
the local quadrupole moments of the radiation distribution.  The temperature
anisotropies and associated intrinsic polarization are well described by the
six-parameter $\Lambda$CDM model, in which dark energy ($\Lambda$) and cold dark
matter (CDM) dominate the energy budget of the universe. This polarization
pattern, also called the $E$-mode, has a distinct curl-free geometric signature
that has been precisely
measured\cite{actpol2014ee,bicep2_spect,bicep_3yr_all,brown2009} and is
consistent with that expected from $\Lambda$CDM.  In fact, the most recent data
from the \planck\ satellite appear remarkably insensitive to any extension
beyond the simple six-parameter case \cite{planck2013_parameters}, such as extra
relativistic degrees of freedom, running of the scalar spectral index, or
spatial curvature.

The next generation of cosmological experiments aim to push the limits of the
$\Lambda$CDM paradigm.  The leading hypothesis for the mechanism that shaped the
nearly flat, homogeneous and isotropic universe that we observe is a period of
super-luminal inflation, occurring in the first instants after the Big Bang.
While these characteristics, together with the nearly scale-invariant spectrum
of primordial fluctuations and the lack of any detectable departures from
Gaussianity provide substantial support for the inflationary paradigm, the
detection of an appreciable level of tensor fluctuations via a gravitationally
sourced quadrupole would represent significant evidence in favor of the most
simple of the inflationary theories. The simplest forms of inflationary models,
typically driven by a single scalar field or effective field, predict a curl or
$B$-mode polarization in the CMB at degree angular scales, due to tensor
fluctuations induced by early universe gravitational waves.\cite{guth2013} The
$B$-mode amplitude scales with the tensor-to-scalar ratio $r$, and is expected
to peak at degree angular scales with an amplitude 6-7 orders of magnitude
fainter than the temperature anisotropy signal.  Other hypotheses, such as the
ekpyrotic scenario\cite{boyle2004}, do not easily admit tensor fluctuations at
detectable levels, so a detection of a $B$-mode signal at large angular scales
would favor inflationary models.  Alternatively, a stringent limit of $r<0.01$
would significantly restrict the allowed range of the inflationary scalar field.
The strongest limit on $r$ to date is $r<0.11$ (95\%~CL), which comes from the
\planck, ACT and SPT temperature power spectra in conjunction with WMAP 9-year
polarization data\cite{planck2013_inflation}.  Due to the dominant cosmic
variance contribution to the limit from temperature data, stronger constraints
on $r$ must come from polarized CMB data, where the tensor contribution is more
significant.

The strongest limit from the $B$-mode spectrum alone is about six times greater
than the temperature limit at $r<0.65$ (95\%~CL), from the BICEP1 3-year
analysis.\cite{bicep_3yr_rot_all} Most recently, the BICEP2 team has claimed a
detection of a cosmological $B$-mode signal corresponding to a tensor-to-scalar
ratio greater than the limit from large scale temperature
anisotropies.\cite{bicep2_spect} At the time of this writing, there is some
evidence that a significant foreground contribution may be bright enough to
account for much, if not all, of this
signal.\cite{mortonsonseljak2014,flauger2014} The BICEP2 data represent a
massive improvement over the state of the art in sensitivity to diffuse
polarization.  Any experiment aiming to establish a primordial origin for
observed $B$-mode power must (1) confirm that the signal's electromagnetic
spectrum is distinct from that of any foregrounds, which requires at least three
frequencies for effective separation; (2) confirm that the angular power
spectrum is consistent with a cosmological signal; and (3) confirm that the
signal is statistically isotropic.  {\itshape No single experiment to date,
  including \spider, has the power to definitively address all of these issues.}
However, as this paper will demonstrate, \spider\ is well-poised to make
progress on each of these fronts.

Section \ref{sec:goals} discusses \spider's science goals and expected
sensitivity on the sky.  Section \ref{sec:inst} provides an overview of the
payload systems and proposed flight strategy.  Sections \ref{sec:cryo} and
\ref{sec:perf} cover the cryogenic and detector performance.  Section
\ref{sec:integ} discusses key aspects of the instrument integration campaign.

\section{Science Goals}\label{sec:goals}

\begin{figure}
\centering
\includegraphics[height=0.2\textheight]{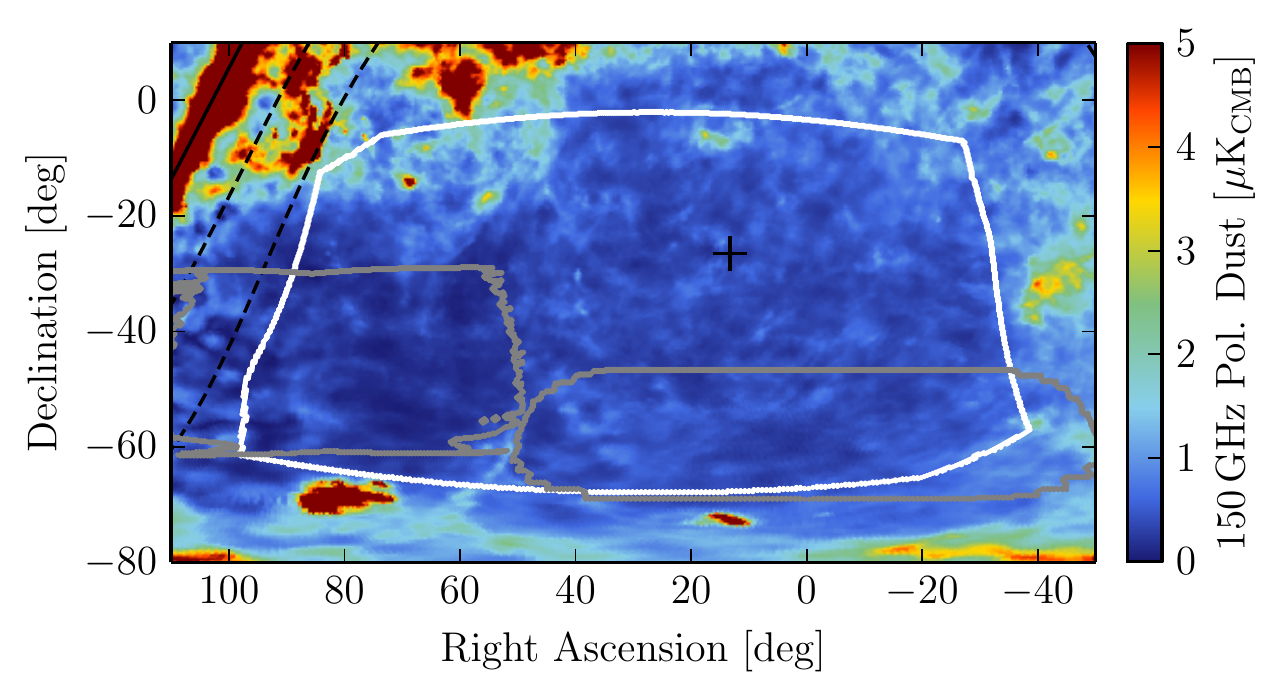}\quad
\includegraphics[height=0.2\textheight]{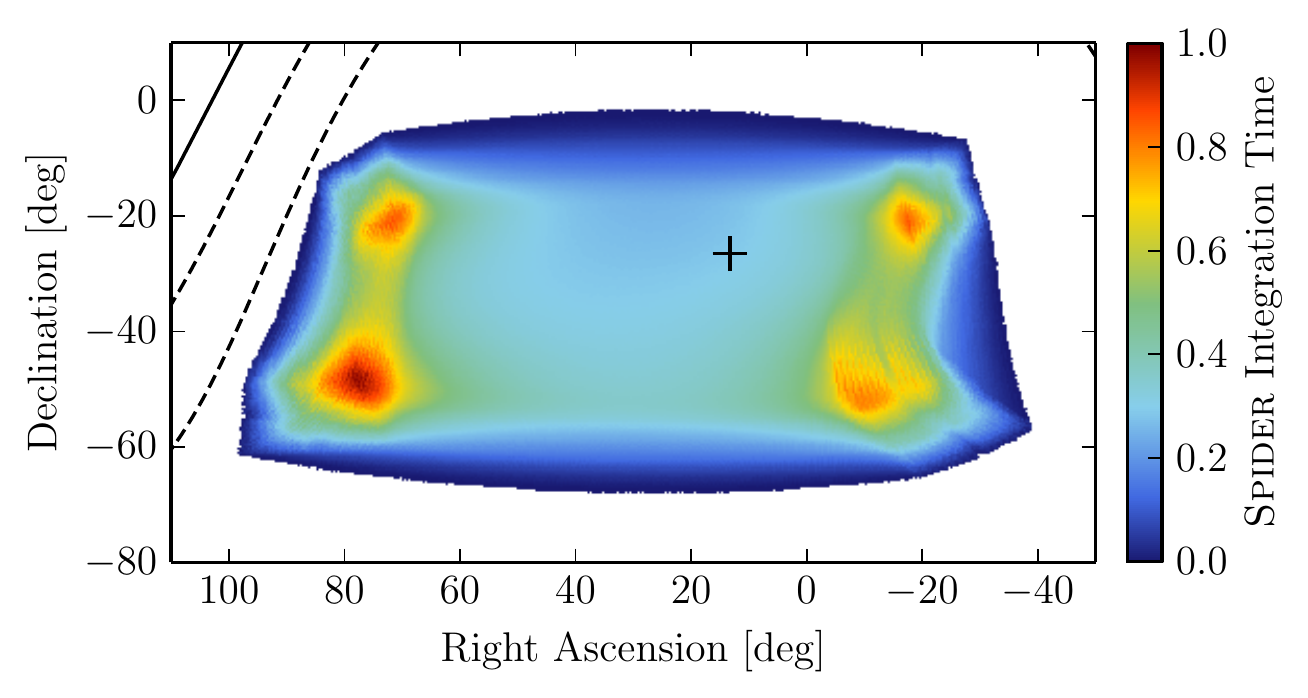}
\caption{({\itshape left}) Amplitude $P=(Q^2+U^2)^{1/2}$ of polarized dust at
  150\,GHz in the southern sky, estimated using the BSS model for dust in the
  Galactic magnetic field\cite{odea2012_dust, fraisse2011_all}.  Galactic
  latitudes $b=0^\circ$ (solid), $b=-10^\circ,-20^\circ$ (dashed), and the
  southern Galactic pole (+) are also shown, along with coverage outlines for
  \spider (white), BOOMERanG (gray, left) and BICEP1/2 (gray, right). ({\itshape
    right}) Distribution of integration time on the sky for a single \spider
  focal plane observing for 24 sidereal hours.}\label{fig:coverage}
\end{figure}

\begin{table}[b]
\centering
\caption{Expected instrument parameters for \spider's first flight in December
  2014, as measured in the lab.  Listed are the telescope center frequencies and
  corresponding bandpass, beam FWHM, total number of light-sensitive detectors,
  detector yield, and per-detector and total sensitivities.  Noise-equivalent
  temperatures (NETs) are estimated from data acquired under flight-like loading
  conditions.  The map depth assumes an effective sky area of 7.5\% covered over
  a 16~day flight with an 85\% duty cycle.}\label{tab:fisher}
\begin{tabularx}{\textwidth}{CCCCCCCC}
\hline\hline
Frequency & Bandpass & FWHM & \multirow{2}{*}{$N_\mathrm{det}$} & Yield & NET$_\mathrm{det}$ & NET$_\mathrm{tot}$ & Depth \\
$[\mathrm{GHz}]$ & [\%] & [arcmin] & & [\%] & [$\mu$K$\sqrt{s}$] & [$\mu$K$\sqrt{s}$] & [$\mu$K$\cdot$arcmin] \\
\hline
94 & 24\% & 42 & 864 & 83\% & 124 & 4.6 & 13.8 \\
150 & 24\% & 30 & 1536 & 85\% & 140 & 3.8 & 11.2 \\
\hline\hline
\end{tabularx}
\end{table}

\spider is a balloon-borne polarized telescope designed to study the microwave
sky at degree angular scales.\cite{spider_exp10_all,fraisse2011_all} We plan to
observe the southern sky at frequencies of 94 and 150\,GHz with an inaugural
flight from McMurdo Station, Antarctica in December 2014.  We prepared and
deployed the completed system for a December 2013 flight, including a successful
integration campaign at NASA's Columbia Scientific Balloon Facility (CSBF) in
Palestine, TX.  However, the U.S. government shutdown resulted in the
cancellation of all flights scheduled for that season.  In the intervening time,
we have been able to perform further characterization and testing of the fully
integrated instrument.  Table~\ref{tab:fisher} shows a summary of the expected
noise performance and map depth after a 16~day flight with an 85\% duty cycle
over 7.5\% of the sky.

The first flight will contain six fully populated receivers, three at each of 94
and 150\,GHz, with each containing 144 and 256 pairs of orthogonal
polarization-sensitive bolometers, respectively.  Each receiver can be
independently aligned at various angles with respect to the scan direction,
allowing simultaneous observation of both $Q$ and $U$ Stokes parameters.
Moreover, an independently controlled stepped half-wave plate (HWP) on each
receiver allows modulation of the sky signal to further improve angular
coverage, and more importantly, mitigate systematic effects that can create a
false polarization signal.  The scan strategy, discussed in more detail in
Section~\ref{sec:scan}, covers the entire observing region once daily.  This
significant level of redundancy and modularity in the instrument allows for many
subdivisions of the data into comparable halves for exhaustive consistency
tests.

By launching off the Antarctic coast, \spider will have access to a relatively
large patch of sky over which to concentrate integration time, as shown in
Figure \ref{fig:coverage}.  The scan pattern allows for observation of 15\% of
the sky while maximizing polarization angle coverage within each pixel;
weighting by integration time to maximize signal-to-noise gives an effective sky
fraction of 7.5\%.  This large coverage area reduces the contribution of sample
variance at the largest scales and permits high fidelity reconstruction of the
shape of the angular power spectrum over many uncorrelated bins at the scales of
interest ($10 < \ell < 100$).  Moreover, a large observing area gives freedom in
selecting independent subregions of the map to test for isotropy of the $B$-mode
signal, and also to enable cross-correlation with other instruments observing
within the same regions of the sky, such as ACTpol\cite{actpol2014ee},
SPTpol\cite{sptpol2012}, BICEP2/Keck\cite{bicep2_spect}, and
\planck\cite{planck2013_overview}.  \spider plans to cover the majority of the
clean sky available from the southern hemisphere.

\begin{figure}
\centering
\includegraphics[width=0.6\textwidth]{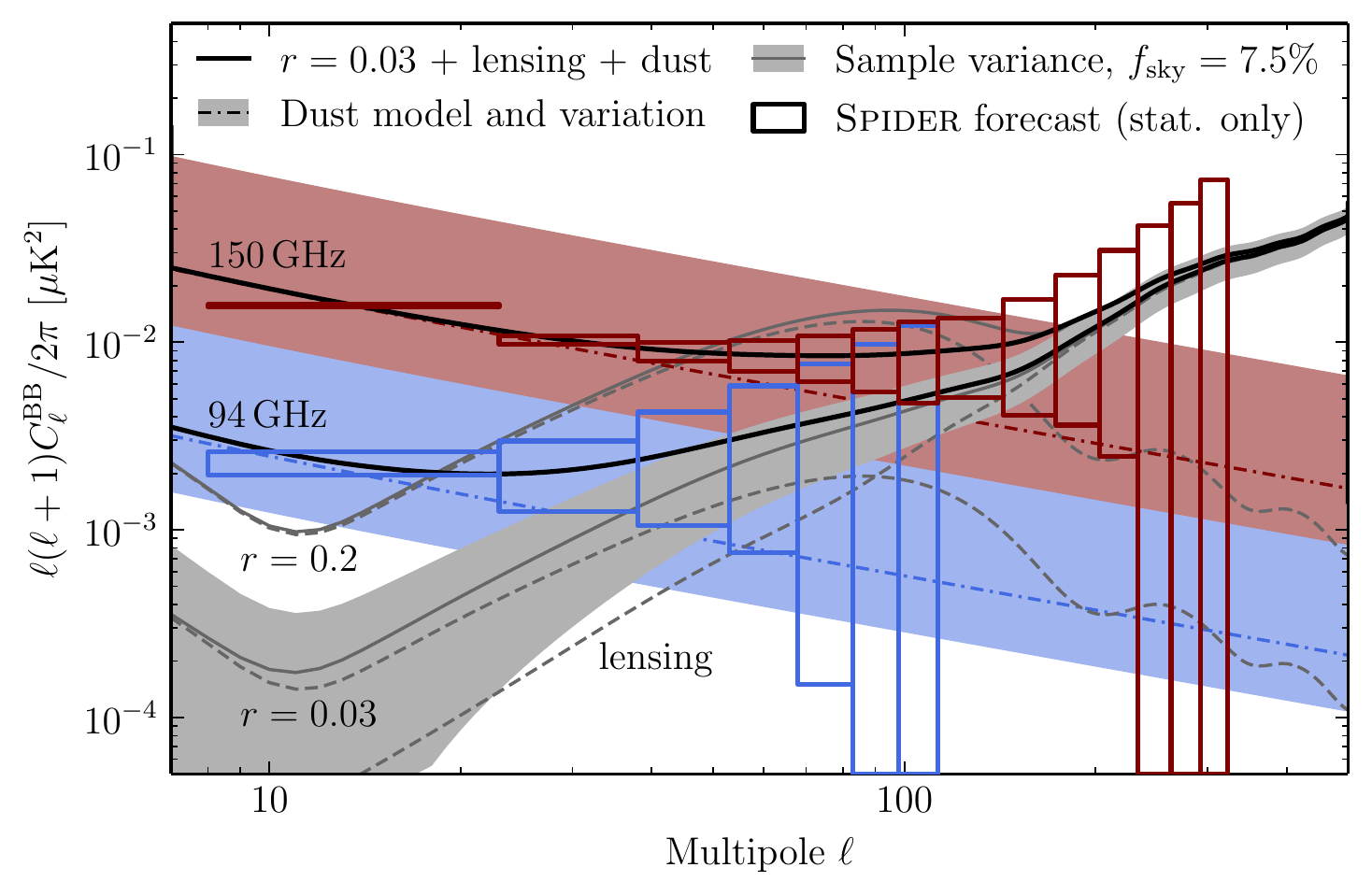}
\caption{Forecast for the December 2014 \spider flight, based on lab
  measurements of representative pixels on each focal plane, and detailed in
  Table~\ref{tab:fisher}.  The expected statistical sensitivity per bin is shown
  independently for the 150\,GHz (red) and 94\,GHz (blue) instruments, overlaid
  on a dust + lensing + $r = 0.03$ angular power spectrum.  The nominal dust
  amplitude at each frequency (dash-dotted line) is that expected over the
  \spider region based on the BSS model of polarized dust
  emission\cite{odea2012_dust, fraisse2011_all}, while the surrounding shaded
  band illustrates the current uncertainty on the dust amplitude in the southern
  Galactic hemisphere; recent data\cite{planck2014_dust,flauger2014} suggest
  that dust temperatures may be as much as a factor of two higher than modeled
  here (corresponding to a factor of four in power).  The cosmological sample
  variance on an $r = 0.03$ + lensing $B$-mode signal observed over 7.5\% of the
  sky is shown in the gray band.  An $r = 0.2$ + lensing spectrum is also shown
  for reference.\label{fig:forecast}}
\end{figure}

By observing at multiple frequencies with similar signal-to-noise, \spider will
also be able to characterize the foreground contribution to the $B$-mode signal.
Figure~\ref{fig:forecast} shows the expected dust contribution to the $B$-mode
power at 94 and 150\,GHz in the \spider observing region, based on the model
shown in the left panel of Figure~\ref{fig:coverage}.  A relatively large band
of uncertainty is also shown, based on the assumption that current models may be
under-estimating the amplitude of polarized dust at high
latitudes.\cite{planck2014_dust,flauger2014} If $r$ is large, \spider will be
able to distinguish the primordial signal from dust at degree scales by
leveraging data from both frequency bands.  If $r$ is relatively small, \spider
should provide a high fidelity map of high-latitude foreground emission at the
CMB frequencies.  A subsequent flight of \spider with instruments at higher
frequencies will then allow foreground separation over the cleanest parts of the
high-latitude sky, and enable detection of primordial signals above the
$r\sim0.03$ level.\cite{fraisse2011_all} Frequency selection and further
optimization of the observing region will be informed by the final release of
polarization data from the \planck team.  The ballooning platform allows access
to a wide range of frequencies without significant atmospheric contamination, so
frequencies higher than those typically accessible from the ground
($\gtrsim$220\,GHz) will be considered.

\section{Instrument Overview}\label{sec:inst}

\subsection{Receiver}

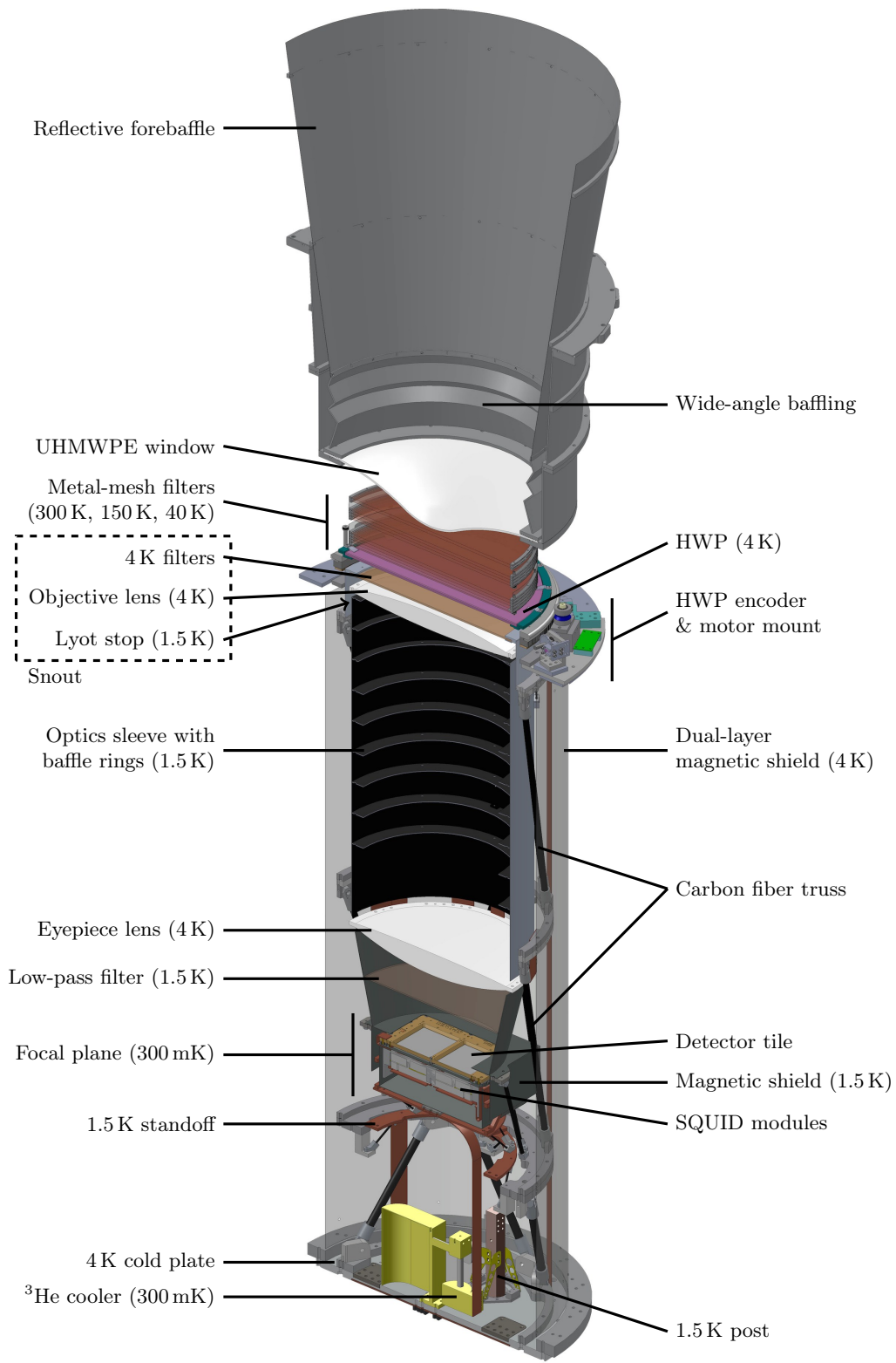
\begin{figure}
\centering
\begin{tikzpicture}
  \tikzstyle{every node}=[font=\small]
  \node[anchor=south west, inner sep=0] (image) at (0,0) {\includegraphics[height=0.95\textheight]{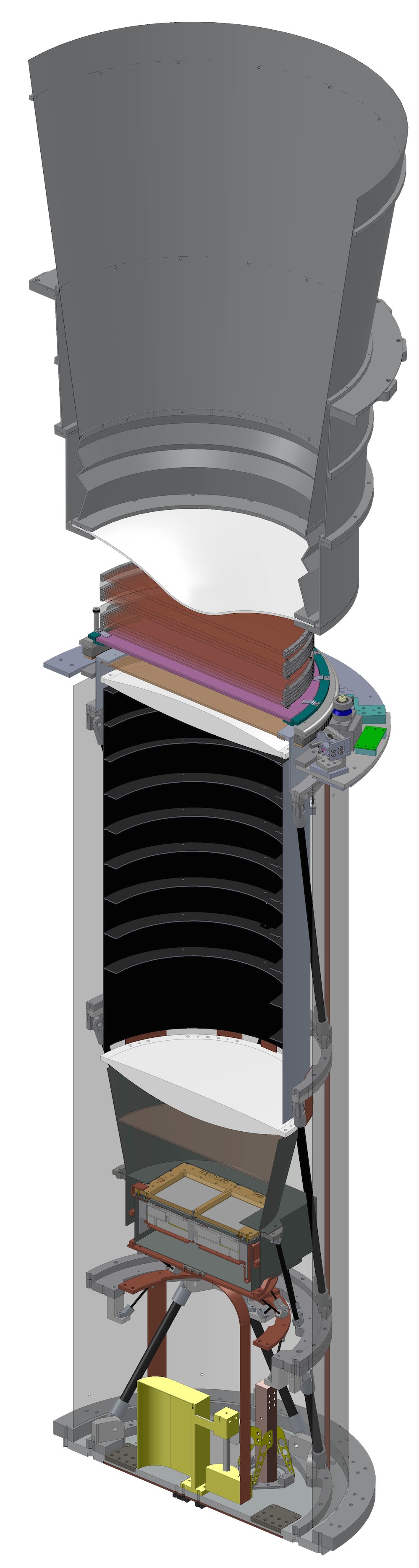}};
  \begin{scope}[x={(image.south east)},y={(image.north west)}]
    \draw[very thick,-] (0.15,0.9) -- (-0.1,0.9);
    \node[anchor=east] at (-0.1,0.9) {Reflective forebaffle};
    \draw[very thick,-] (0.6,0.7) -- (1.1,0.7);
    \node[anchor=west] at (1.1,0.7) {Wide-angle baffling};
    \draw[very thick,-] (0.32,0.65) -- (-0.1,0.67);
    \node[anchor=east] at (-0.1,0.67) {UHMWPE window};
    \draw[very thick,-] (0.18,0.615) -- (-0.1,0.63);
    \draw[very thick,-] (0.18,0.595) -- (0.18,0.635);
    \node[anchor=east,align=right] at (-0.1,0.63) {Metal-mesh filters \\ (300\,K, 150\,K, 40\,K)};
    \draw[very thick,-] (0.71,0.55) -- (1.1,0.6);
    \node[anchor=west] at (1.1,0.6) {HWP (4\,K)};
    \draw[very thick,-] (0.95,0.53) -- (1.1,0.55);
    \draw[very thick,-] (0.95,0.5) -- (0.95,0.56);
    \node[anchor=west,align=left] at (1.1,0.55) {HWP encoder \\ \& motor mount};
    \draw[very thick,-] (0.82,0.45) -- (1.1,0.45);
    \node[anchor=west,align=left] at (1.1,0.45) {Dual-layer \\ magnetic shield (4\,K)};
    \draw[very thick,-] (0.3,0.575) -- (-0.1,0.59);
    \node[anchor=east] at (-0.1,0.59) {4\,K filters};
    \draw[very thick,-] (0.29,0.565) -- (-0.1,0.56);
    \node[anchor=east] at (-0.1,0.56) {Objective lens (4\,K)};
    \draw[very thick,<-] (0.24,0.557) -- (-0.1,0.53);
    \node[anchor=east] at (-0.1,0.53) {Lyot stop (1.5\,K)};
    \draw[very thick, dashed, -] (-0.66,0.515) rectangle (-0.08,0.605);
    \node[anchor=north west] at (-0.65,0.515) {Snout};
    \draw[very thick,-] (0.28,0.45) -- (-0.1,0.45);
    \node[anchor=east,align=right] at (-0.1,0.45) {Optics sleeve with \\ baffle rings (1.5\,K)};
    \draw[very thick,-] (0.3,0.32) -- (-0.1,0.32);
    \node[anchor=east] at (-0.1,0.32) {Eyepiece lens (4\,K)};
    \draw[very thick,-] (0.32,0.285) -- (-0.1,0.285);
    \node[anchor=east] at (-0.1,0.285) {Low-pass filter (1.5\,K)};
    \draw[very thick,-] (0.57,0.23) -- (1.1,0.24);
    \node[anchor=west] at (1.1,0.24) {Detector tile};
    \draw[very thick,-] (0.7,0.21) -- (1.1,0.21);
    \node[anchor=west] at (1.1,0.21) {Magnetic shield (1.5\,K)};
    \draw[very thick,-] (0.54,0.205) -- (1.1,0.18);
    \node[anchor=west] at (1.1,0.18) {SQUID modules};
    \draw[very thick,-] (0.25,0.23) -- (-0.1,0.23);
    \draw[very thick,-] (0.25,0.2) -- (0.25,0.26);
    \node[anchor=east] at (-0.1,0.23) {Focal plane (300\,mK)};
    \draw[very thick,-] (0.31,0.18) -- (-0.1,0.18);
    \node[anchor=east] at (-0.1,0.18) {1.5\,K standoff};
    \draw[very thick,-] (0.75,0.38) -- (1.1,0.35);
    \draw[very thick,-] (0.73,0.25) -- (1.1,0.35);
    \node[anchor=west] at (1.1,0.35) {Carbon fiber truss};
    \draw[very thick,-] (0.53,0.055) -- (-0.1,0.055);
    \node[anchor=east] at (-0.1,0.055) {$^3$He cooler (300\,mK)};
    \draw[very thick,-] (0.2,0.08) -- (-0.1,0.08);
    \node[anchor=east] at (-0.1,0.08) {4\,K cold plate};
    \draw[very thick,-] (0.64,0.07) -- (1.1,0.03);
    \node[anchor=west] at (1.1,0.03) {1.5\,K post};
  \end{scope}
\end{tikzpicture}

\caption{A section view of the receiver, showing the refractive optics, filters,
  baffling, and focal plane assembly.}\label{fig:insert}
\end{figure}

\spider was designed to be a relatively modular instrument, with room for six
independently controlled receivers, one of which is shown in
Figure~\ref{fig:insert}.  The design of the receivers is discussed in more
detail in Runyan (2010)\cite{spider_instr10_all}.  The optical system is a
simple refractor, as originally designed for the the BICEP1
instrument\cite{bicep1_instr}.  The aperture of each receiver is 25\,cm with a
20$^\circ$ field of view, optimized for degree-scale resolution on the sky.  The
window for each aperture is a relatively thin (1/8'') ultra-high molecular
weight polyethylene (UHMWPE) sheet, anti-reflection (AR) coated for the
appropriate frequency band.  The window is mounted in a reentrant ``bucket''
which accommodates a lightweight reflective baffle protecting the telescope from
stray rays outside of its 20$^\circ$ field-of-view.  A carbon-fiber structure
provides rigid support of the focal plane and lenses with minimal coefficient of
thermal contraction, to avoid distortion of the beam.  The high-density
poylethylene (HDPE) lenses are simple conics in a telecentric configuration to
allow a straightforward mapping of the flat focal plane onto the sky, and are
cooled to 4\,K to reduce in-band loading due to dielectric loss.  A blackened
sleeve placed between the two lenses is cooled to 1.5\,K to further reduce
instrument loading, and also provides a beam-defining Lyot stop just below the
objective lens.  Infrared loading is attenuated by an AR-coated 10\,\icm
metal-mesh filter and an AR-coated 1/8'' nylon dielectric filter, placed in the
``snout'' above the objective.  An independently-controlled sapphire half-wave
plate (HWP) is mounted above the snout on each receiver for modulating
polarization on the sky.\cite{spider_hwp10_all} An Amuneal\circledR\ Amumetal-4K
magnetic shield cooled to 1.5\,K surrounds the focal plane and supports a final
low-pass metal-mesh filter (4\,\icm and 6\,\icm for 94 and 150\,GHz,
respectively) and above the detectors.  The AR-coatings and filters are all
individually optimized for the \spider frequency bands.  The focal plane is
stood off from the 1.5\,K stage using a smaller carbon-fiber structure, and
cooled to 300\,mK using a closed-cycle helium-3 adsorption refrigerator.
Stainless steel passive thermal filters reduce responsivity to thermal
fluctuations in the fridge, with a time constant of about 80 seconds.  Flexible
copper heat straps provide thermal contact to the refrigerator condensation
point (1.5\,K) and cold head (300\,mK), while solid copper bus bars along the
length of the receiver maintain a low thermal gradient between the cold plate
and the snout.

\subsection{Focal Plane}

\begin{figure}
\centering
\includegraphics[height=0.18\textheight]{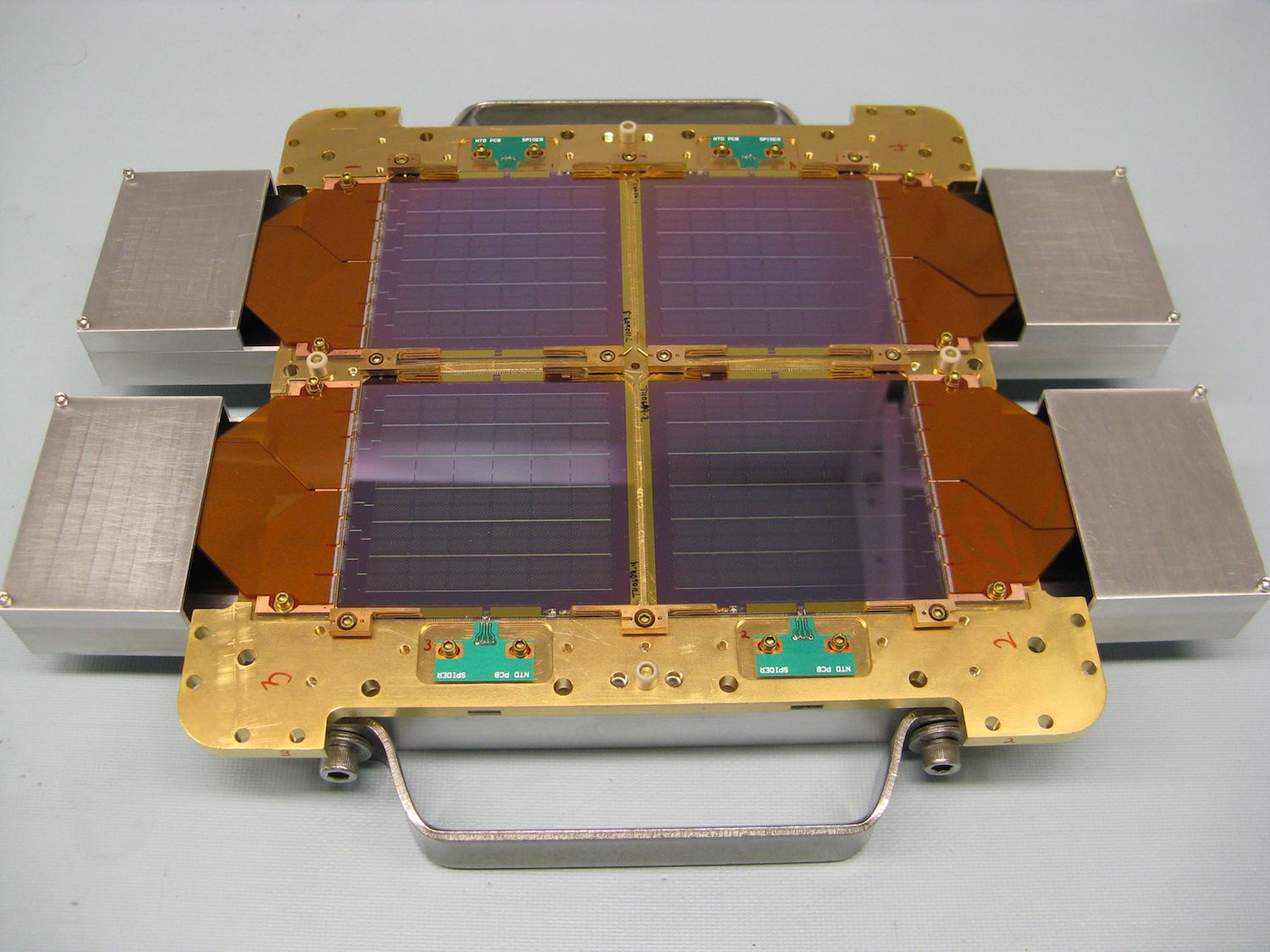}\hfill
\begin{tikzpicture}
  \tikzstyle{every node}=[font=\sffamily\footnotesize]
  \node[anchor=south west, inner sep=0] (image) at (0,0) {\includegraphics[height=0.18\textheight]{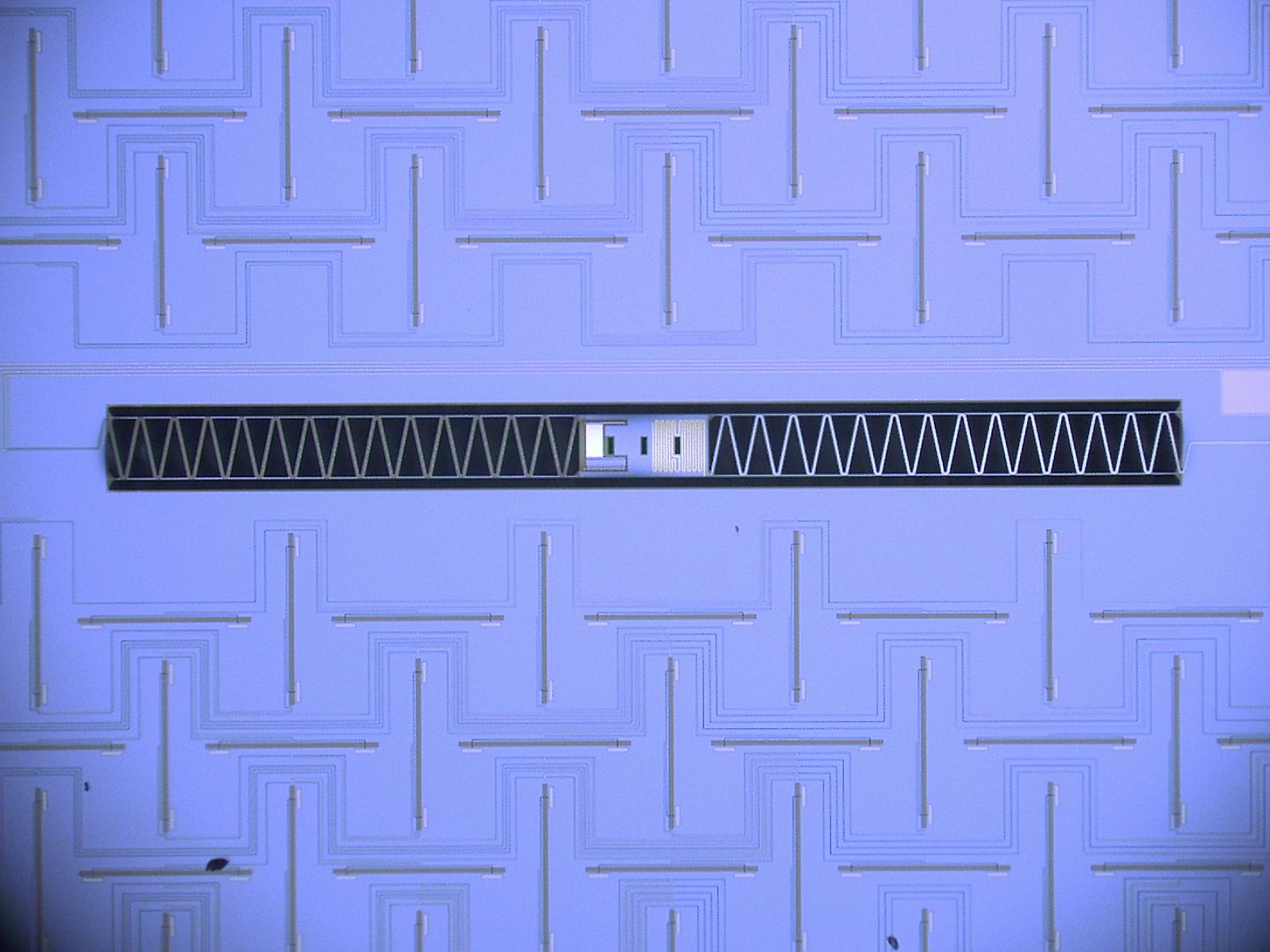}};
  \begin{scope}[x={(image.south east)},y={(image.north west)}]
    \draw[white,very thick] (0.38,0.4) rectangle (0.635,0.66);
    \draw[white,very thick,<->] (0.63,0.095) -- (0.82,0.095);
    \node[white,anchor=south] at (0.725,0.11) {0.5\,mm};
  \end{scope}
\end{tikzpicture}
\hfill
\begin{tikzpicture}
  \tikzstyle{every node}=[font=\sffamily\small]
  \node[anchor=south west,inner sep=0] (image) at (0,0) {\includegraphics[height=0.18\textheight]{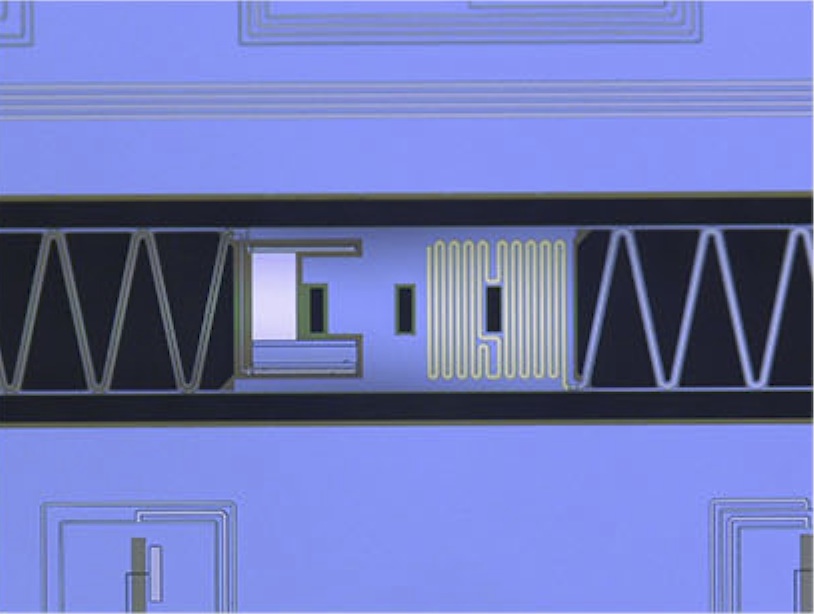}};
  \begin{scope}[x={(image.south east)},y={(image.north west)}]
    \node[white] at (0.36,0.25) {Ti TES};
    \draw[white, very thick,->] (0.34,0.3) -- (0.34,0.4);
    \node[white] at (0.36,0.74) {Al TES};
    \draw[white, very thick,->] (0.34,0.69) -- (0.34,0.58);
    \node[white] at (0.62,0.25) {Resistor};
    \draw[white, very thick, ->] (0.6,0.3) -- (0.6,0.4);
    \node[white,anchor=east] at (0.85,0.74) {Antenna};
    \draw [white,very thick,->] (0.85,0.74) -- (0.95,0.74);
  \end{scope}
\end{tikzpicture}
\caption{(\textit{left}) The underside of a fully populated \spider 150\,GHz
  focal plane during assembly, with the $8 \times 8$ grid of pixels visible on
  each of the four tiles.  (\textit{middle}) A microscope image of a bolometer
  island and surrounding dipole antenna array.  (\textit{right}) A close-up
  image of the island, with both TES devices and the meandering gold resistor
  clearly visible.}\label{fig:fpu}
\end{figure}
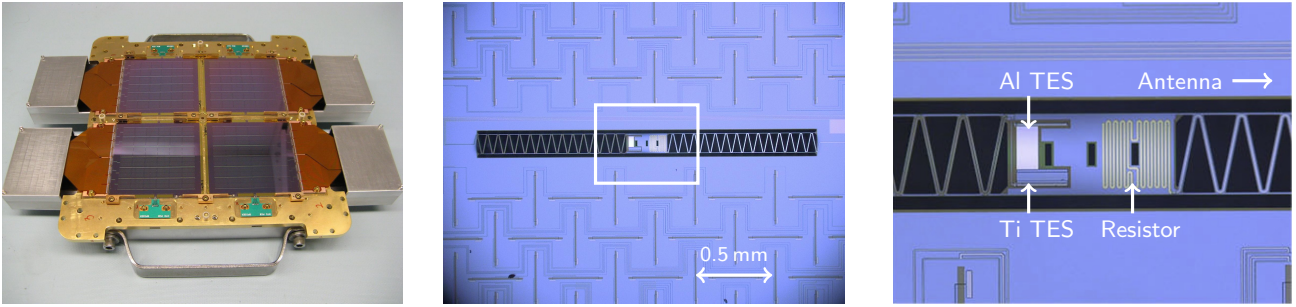

The focal plane and cold readout electronics have been optimized for a sensitive
polarimetric measurement at millimeter wavelengths.\cite{obrient2014} Both the
beam-forming and bolometric elements of the detectors are lithographed in a grid
pattern onto an SiN wafer, as shown in Figure~\ref{fig:fpu}.  Each physical
pixel consists of two detectors, one of which is sensitive to vertical
polarization, and the other of which is sensitive to horizontal polarization.
The detector beams are formed by an interleaved array of horizontal and vertical
slot dipole antennas, with each polarization fed via microstrip lines to
independent bolometer islands.  The bolometers are superconducting
transition-edge sensors (TES) with thermal conductivities tuned for
balloon-borne observations.  Each bolometer island contains a meandering gold
resistor onto which power from the antennas is dissipated, a titanium TES ($T_c
\sim 500$\,mK, $R_n \sim 30$\,m$\Omega$) with a typical saturation power of
$\sim$2-3\,pW for science observations in flight, as well as an aluminum TES
($T_c \sim 1.1$\,K, $R_n \sim 0.1\,\Omega$) with a saturation power of
$\sim$80\,pW which allows ground-based end-to-end testing of the optical systems
using LN$_2$ or room-temperature sources.

The detectors are read out using a three-stage superconducting quantum
interference device (SQUID) system with time-domain multiplexing
electronics.\cite{battistelli2008} The layout of the focal plane and SQUID
modules was designed to accommodate a multi-layer magnetic shielding
architecture, necessary for \spider's wide-area scan and motion within the
Earth's varying magnetic field.  The expected shielding factor of this design is
$10^8$, an order of magnitude better than necessary to maintain stable detector
operation.\cite{spider_instr10_all}

\subsection{Cryogenic System}

\begin{figure}
\centering
\includegraphics[height=0.3\textheight]{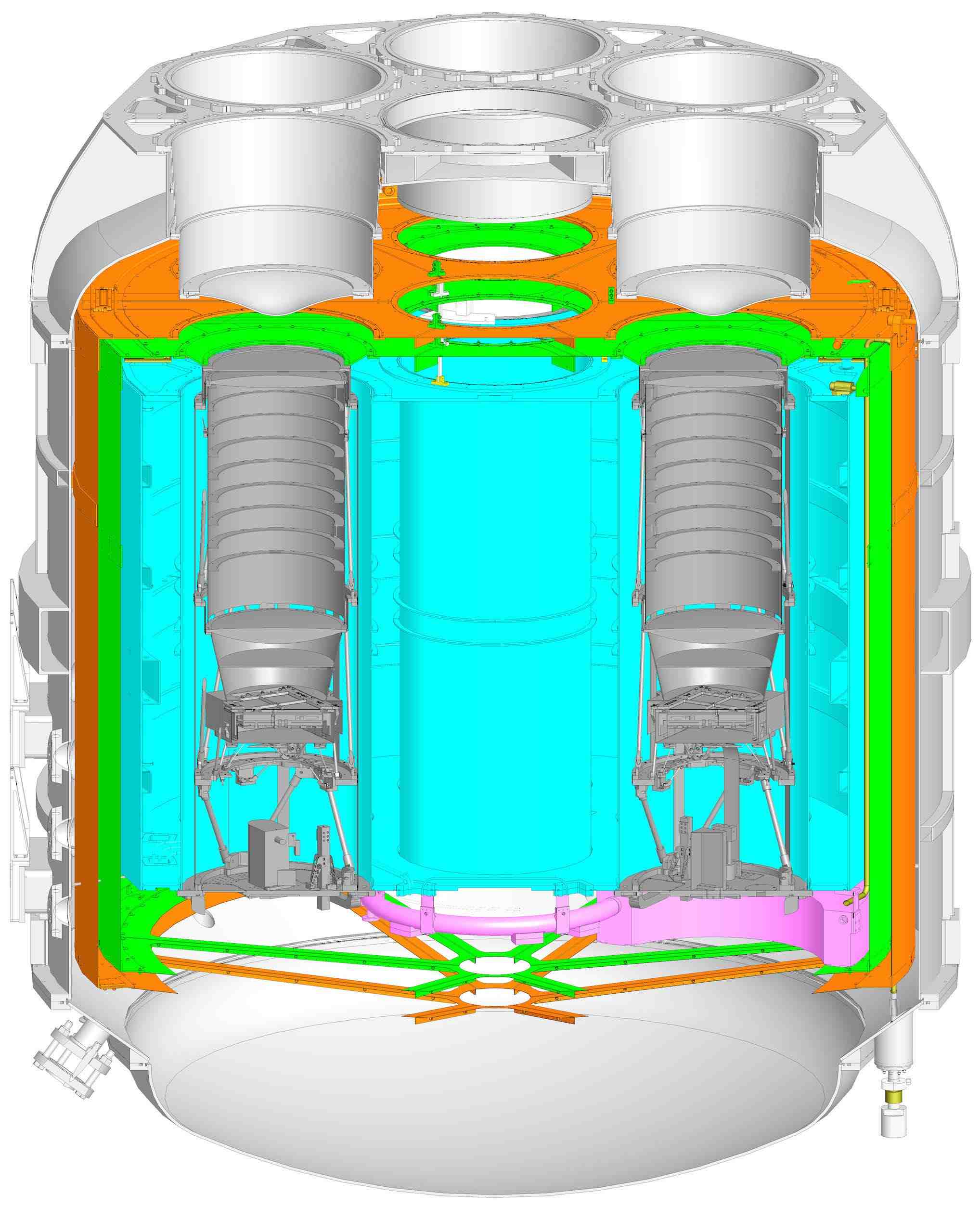}\qquad\qquad
\includegraphics[height=0.3\textheight]{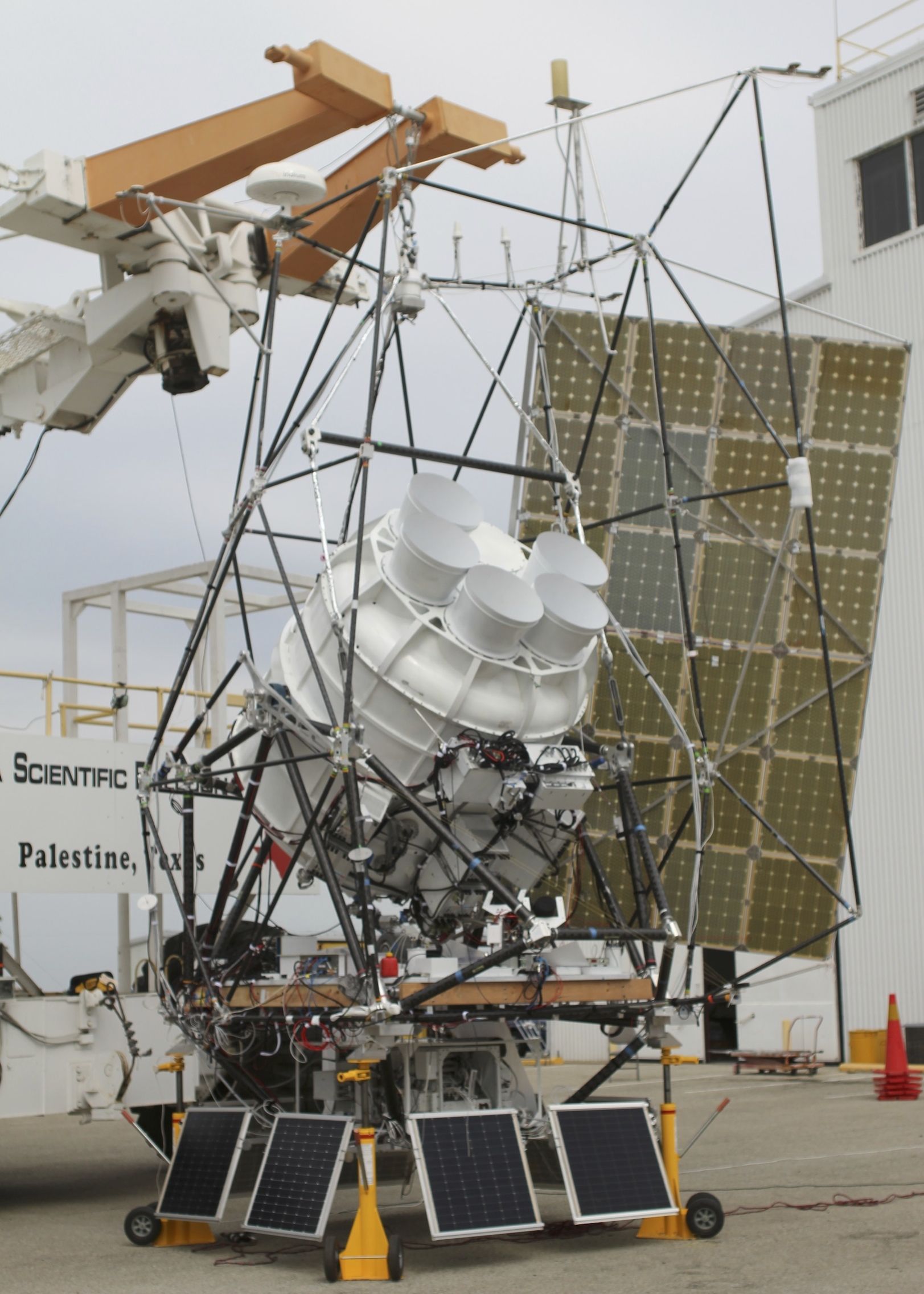}
\caption{(\textit{left}) The cryostat with each temperature stage highlighted:
  VCS2 (orange), VCS1 (green), main tank (cyan), and superfluid tank (pink).
  The receivers (dark gray) are shown in detail in Figure
  \ref{fig:insert}. (\textit{right}) The cryostat mounted on the fully assembled
  gondola during the 2013 integration campaign.}\label{fig:cryo}
\end{figure}

The \spider receivers are housed in a large (2\,m diameter) liquid helium
cryostat\cite{spider_cryo10_all}, shown in Figure \ref{fig:cryo}.  The main tank (hereafter MT) can hold up to 1300\,L liquid He, and provides
4\,K cooling power to each receiver via flexible copper straps in addition to a
thermal bus running internally to the cryostat.  The main tank also supplies
liquid to the pumped auxiliary superfluid tank (hereafter SFT) via a
custom-designed capillary system (Section \ref{sec:capillary}).  The SFT provides a
1.5\,K condensation point for each of the closed-cycle helium-3 adsorption
refrigerators inside each receiver.  The boil-off vapor from the main tank also
provides cooling power to two vapor-cooled shielding stages (VCS1 and VCS2) via
a series of heat exchangers.  Multi-layer insulation (MLI) between the vacuum
vessel wall and each of the VCS stages reduce radiative loading on the colder
stages.  The receiver apertures at each VCS stage are covered with a combination
of single and multilayer reflective low pass metal-mesh
filters,\cite{tucker2006,ade2006} designed to minimize infrared loading on the
cryogenic system, while maintaining adequate microwave throughput to the
detectors (Section \ref{sec:radfilt}).  This system is designed to provide more
than 20 days of hold time under flight loading conditions.  Temperature control
and monitoring is provided by custom housekeeping electronics.\cite{benton2014}

\subsection{Gondola and Pointing Systems}

The cryostat is supported on a light-weight carbon fiber
gondola\cite{soler2014mecha}, along with the flight electronics, pointing
sensors and motors, and communications equipment.  The gondola is able to point
the receivers using a combination of a reaction wheel and pivot for azimuth
control and a pair of linear actuators for elevation control\cite{jas2014}.  The
pointing system was designed to scan in azimuth with a sinusoidal velocity
profile, providing speeds of up to 6\,deg/s, and accelerations up to
0.8\,deg/s$^2$.  A variety of sensors are used to determine the in-flight and
post-flight pointing solution\cite{gandilo2014}.  A differential GPS, a
magnetometer, pin-hole sun sensors, clinometers and elevation encoders are used
to reconstruct the in-flight pointing solution to an accuracy of a few minutes
of arc.  Post-flight pointing reconstruction is done using images from an
orthogonal pair of tracking star cameras, a boresight star camera, and a 3-axis
gyroscope, with an ultimate accuracy of about 3\,arcsec.  The monitoring and
control of the pointing and cryogenic systems are done using custom electronics
and software built on the legacy of the BLAST and BLASTpol control
systems.\cite{blast_inst,benton2014}

\subsection{Scan Strategy}\label{sec:scan}

\begin{figure}
\centering
\includegraphics[height=0.195\textheight]{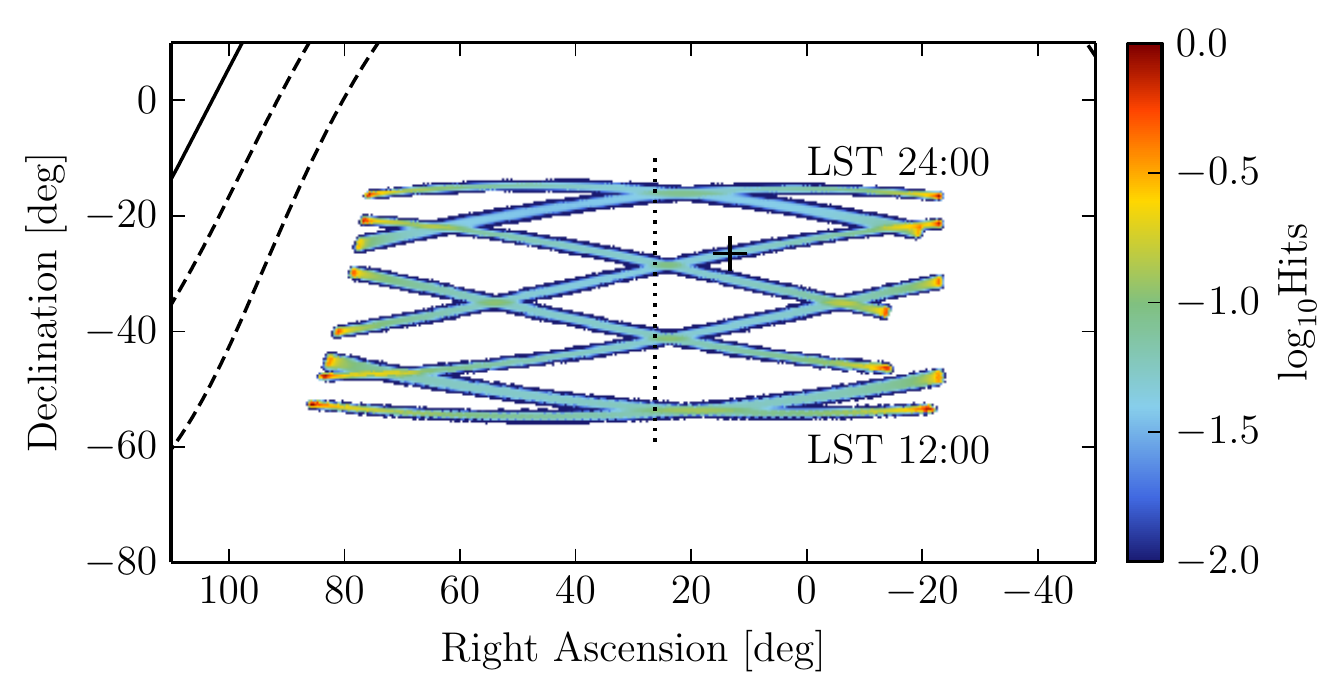}\quad
\includegraphics[height=0.195\textheight]{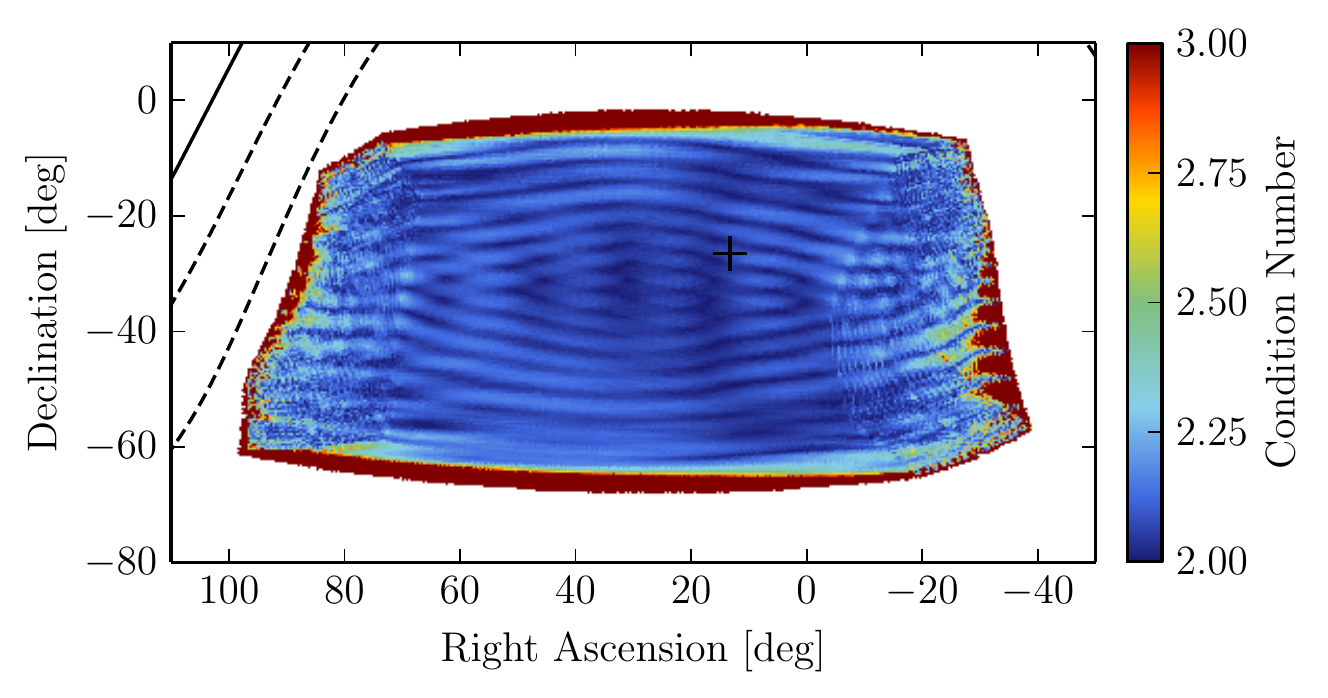}
\caption{({\itshape left}) The scan pattern for a single detector on the sky
  over the course of one sidereal day.  A five-minute period every three hours
  is shown.  Each back-and-forth scan period in azimuth goes through a track
  point at the center of the scan region (dotted line).  The track point is
  adjusted for uniform coverage in declination over a twelve-hour period.  The
  lowest track point declination is achieved near sidereal noon, and the highest
  near sidereal midnight.  Sky rotation provides good cross-linking between the
  upward- and downward-moving half-day scans.  ({\itshape right}) Combined with
  22.5$^\circ$ steps of the HWP every 12 sidereal hours, uniform polarization
  angle coverage is achieved throughout the map, as evidenced by the condition
  number of the pointing matrix for a full 256-pixel/512-detector focal plane.
  The condition number is 2 in the limit of ideal angular coverage. Galactic
  latitudes $b=0^\circ$ (solid), $b=-10^\circ,-20^\circ$ (dashed), and the
  southern Galactic pole (+) are also shown in each panel.}\label{fig:scan}
\end{figure}

The nominal \spider scan strategy was designed to maximize coverage over the
largest amount of accessible clean sky in the southern hemisphere.  The scan
region is limited in right ascension by proximity to the Galaxy or the sun, and
in declination by the payload's latitude and the extension of the elevation
arms.  A sinusoidal scan profile was chosen to limit the amount of torque that
the gondola and flight train must exert.\cite{jas2014} The boresight motion
across the scan region at different times of the sidereal day is shown in
Figure~\ref{fig:scan}.  Later in the summer season (late December to January)
the sun moves closer to the Galaxy and restricts the available sky with low
foreground contamination, so an early launch is favorable for \spider. Rotation
of the sky within the field of view provides good angular cross-linking across
the extent of the map, as long as the payload remains at latitudes of about
$-80^\circ$ or higher as it follows the circumpolar winds around the continent.
To further improve angular coverage, the HWP is stepped by 22.5$^\circ$ every 12
sidereal hours.

For reconstruction of the $I$, $Q$ and $U$ Stokes parameters from the detector
time streams, a set of linear equations is constructed for each map
pixel:\cite{jones2007}
\begin{equation}
\left(\begin{array}{c}
\left< d \right> \\
\left< d \gamma \cos 2\psi \right> \\
\left< d \gamma \sin 2\psi \right>
\end{array}\right) = 
\left(\begin{array}{ccc}
1 & \left< \gamma \cos 2\psi \right> & \left< \gamma \sin 2\psi \right> \\
\left< \gamma \cos 2\psi \right> & \left< \gamma^2 \cos^2 2\psi \right> & \left< \gamma^2 \cos 2\psi \sin 2\psi \right> \\
\left< \gamma \sin 2\psi \right> & \left< \gamma^2 \cos 2\psi \sin 2\psi \right> & \left< \gamma^2 \sin^2 2\psi \right> \\
\end{array}\right)
\left(\begin{array}{c}
I \\
Q \\
U \\
\end{array}\right),
\label{eq:inv}
\end{equation}
where the data $d$ are noisy observations of the microwave sky,
$\left<\ldots\right>$ indicates an average over time stream elements falling
into the map pixel, $\gamma$ is the polarization efficiency of the given
channel, and $\psi = \phi + 2\theta$ is the projected polarization angle onto
the map pixel, with detector angle $\phi$ and additional rotation 2$\theta$
provided by the HWP.  Many methods have been developed for optimal inversion of
these equations.  Additionally, the inversion becomes more complicated if
non-idealities in the HWP must be taken into account\cite{bmr2010}.  However, we
can make a simple estimate of how well the Stokes parameters can be
reconstructed by estimating the condition number of the matrix in
equation~\ref{eq:inv} (the ratio of the largest and smallest singular values of
the matrix), shown in Figure~\ref{fig:scan} and averaged over all detectors on a
focal plane over a single sidereal day of observation, in the ideal scenario of
an ideal HWP and $\gamma = 1$.  In the limit of perfect angular coverage,
i.e. if both the $Q$ and $U$ Stokes parameters are measured with equal
sensitivity, the condition number should approach 2; thus, the scan strategy as
designed is quite uniform and near ideal.

The scan strategy naturally divides a single day's observation into two halves:
the upward-moving scan before sidereal midnight, and the downward-moving scan
after sidereal midnight.  The same scan is repeated each day, with a different
pair of HWP angles during each half of the day.  Observation over eight days, at
all eight unique HWP angles during each half day, provides an optimal set of
maps from which to correct for systematic effects, most notably beam systematics
or ghosting, by taking appropriate linear combinations of half-day maps to
create templates or subtract out these effects.\cite{sab_thesis}

\begin{table}
\centering
\caption{A selection of null tests that can be performed on the \spider dataset
  to probe various systematic effects.}\label{tab:jacks}
\begin{tabular}{ll}
\hline\hline
Null Test & Systematics Probed \\
\hline
Upward-moving vs downward-moving & transfer function, gain drifts, ground pickup \\
Left-moving vs right-moving & transfer function, galactic sidelobes \\
Inner pixels vs outer pixels & beam ellipticity, ghosting, sidelobes \\
A pol vs B pol & cross-polar leakage \\
Left tiles vs right tiles & sidelobes \\
Even rows vs odd rows & electrical cross-talk \\
94\,GHz vs 150\,GHz & foregrounds \\
Left receiver vs right receiver & gain drifts, sidelobes, polarization offsets \\
Early days vs late days & gain drifts \\ 
\hline\hline
\end{tabular}
\end{table}

The modularity of the instrument and scan strategy lend itself very
naturally to a large selection of null tests to probe systematic effects.  An
angular power spectrum of the difference between two halves of the dataset, when
compared against simulations of expected variation in the signal and noise,
should yield a spectrum consistent with a null result, if all instrumental and
systematic effects are well understood.  Table \ref{tab:jacks} lists a selection
of such consistency tests.  This is by no means an exhaustive list, but is meant
to illustrate the discriminatory power of the \spider dataset.

\section{Cryogenic Performance}\label{sec:cryo}

\subsection{Equilibrium Performance and Hold Time}

The SPIDER cryostat's main tank (MT) can hold up to 1300 L of liquid helium.  It
was designed to stay cold throughout a long duration balloon flight, which
typically last from 10 to 20 days.  Recent testing with all six receiver ports
populated indicates equilibrium temperatures of 40\,K and 150\,K at VCS1 and
VCS2, respectively.  The equilibrium gaseous flow rate is $\sim$40 SLPM through
the VCS heat exchangers, corresponding to a load of about 2.5\,W. This indicates
an expected hold time of roughly 16 days.  Prior testing with one or two
populated receivers indicated a 40\% lower load on the MT.  The reduced hold
time is thought to be due in part to increased radiative loading through the
apertures, discussed in more detail in the following section.

\subsection{Radiative Filtering}\label{sec:radfilt}

The \spider filter stack was designed to be effective under ground loading
conditions, where the incident radiation is typically a 300\,K blackbody peaking
at $\sim$16\,THz, and float loading conditions, where the incident radiation is
the 2.7\,K CMB blackbody above an emissive atmosphere.  A combination of thick
dielectric filters, thin metal-mesh reflective shaders,\cite{tucker2006} and
multi-layer hot-pressed metal-mesh filters\cite{ade2006} are used to achieve the
desired in-band loading on the detectors and infrared loading on the system.  A
detailed model of the transmission and absorption of these filters was used to
inform modifications for improving cryogenic performance.

\subsubsection{Window}

The vacuum is maintained by a \sfrac{1}{8}'' ultra-high molecular weight
polyethylene (UHMWPE) window, with an anti-reflection coating (Porex\circledR\
PM23DR) applied to both sides.  An excess load has been observed on the
cryogenic system, some of which could be due to the emissive window.  Because
the potential for light leaks down the optical path is high, cutting off the
load from the window before it reaches the colder stages is expected to improve
cryogenic performance.  To reduce infrared loading on the VCS stages from the
window, a pair of baffled shaders have been installed directly below the window;
the effect of these shaders on cryogenic performance will be evaluated in
upcoming testing prior to deployment.  Near the detector passbands, the window
is expected to contribute about 0.1\,pW to the internal load on the detectors.

\subsubsection{VCS1 and VCS2}

The VCS2 temperature stage typically reaches about 150-160\,K, which is too
hot to support any absorptive filters, so only single-layer IR shaders are
installed.  Four shaders provide a $1$\,THz cut-off with 90\% passband
transmission and $< 5\%$ stopband transmission.  The load onto VCS2 is expected
to be about 2\,W per aperture, and transmit about 1\% of the incident radiation
on to VCS1.

The VCS1 stage houses three additional shaders and a 12\,\icm multi-layer
filter.  To further attenuate radiation above 1\,THz, an AR-coated nylon filter
was considered; however at the nominal $\sim$40\,K temperature of the VCS1
stage, modeling of the nylon filter suggested it could be emitting as much as
70\,mW ($\sim$400\,mW for all six receivers), which represents a substantial
fraction of the load onto the main tank; removal of this filter was seen to
improve cryogenic performance in recent testing.

\subsubsection{Receiver Filters}

Additional filters are installed within each receiver assembly.  A 10\,\icm
AR-coated multi-layer low-pass filter is the top-most 4\,K filter to reflect the
majority of infrared radiation.  An AR-coated nylon filter is placed just below
that to attenuate any blue leaks above 1\,THz.  A 6\,\icm low-pass filter
(4\,\icm in the 94\,GHz receivers), installed within the 2\,K magnetic shield,
defines the upper edge of the detector passband.  The total in-band loading onto
the detector has been measured in a test system to be about
0.5\,pW,\cite{art_thesis} consistent with expectations from modeling of the
filter stack.  Additional baffling, discussed in detail in Section
\ref{sec:baffle}, should significantly limit excess loading beyond this.

\subsection{Superfluid Capillary}\label{sec:capillary}

\begin{figure}
\centering
\includegraphics[height=0.25\textheight]{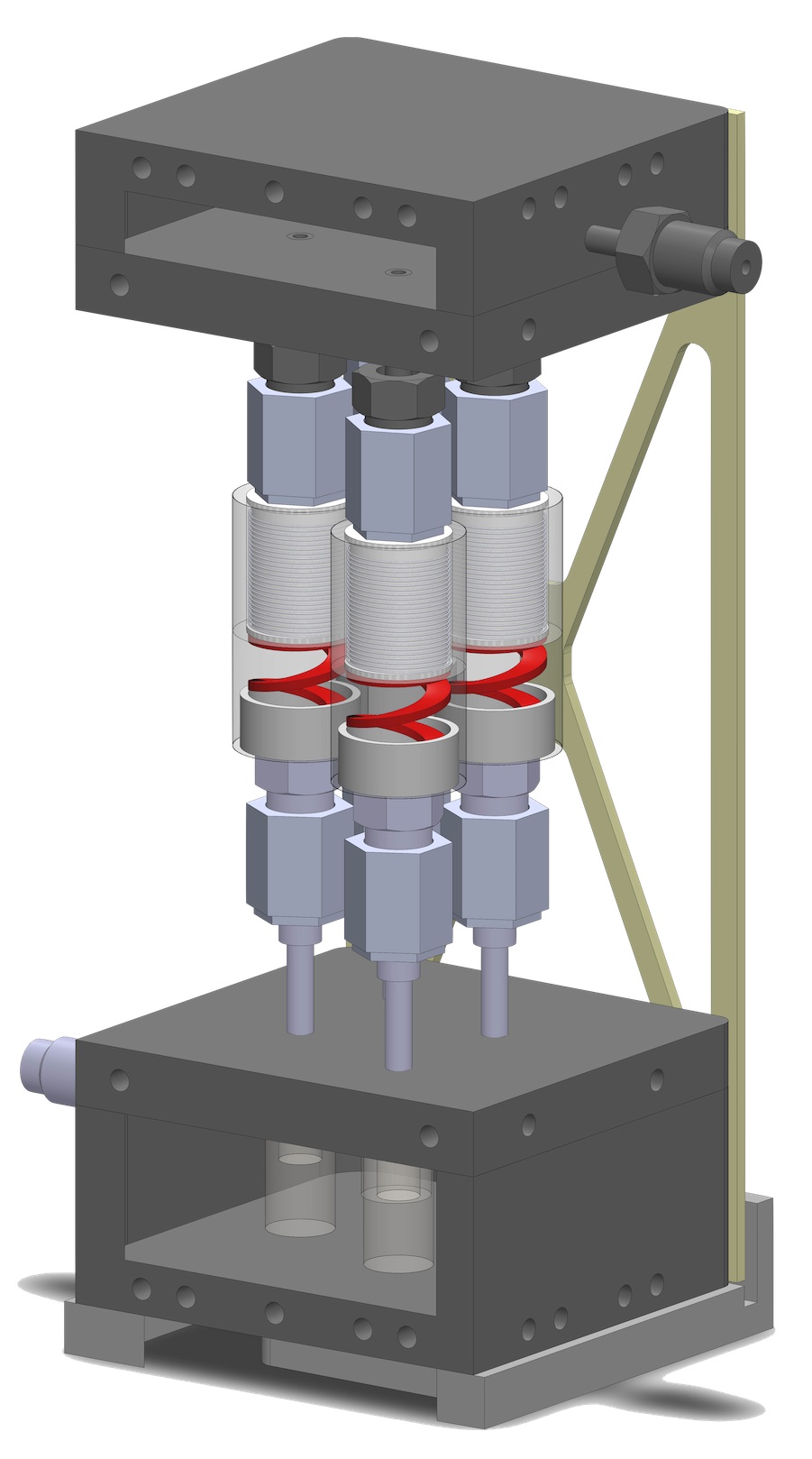}\qquad
\includegraphics[height=0.25\textheight]{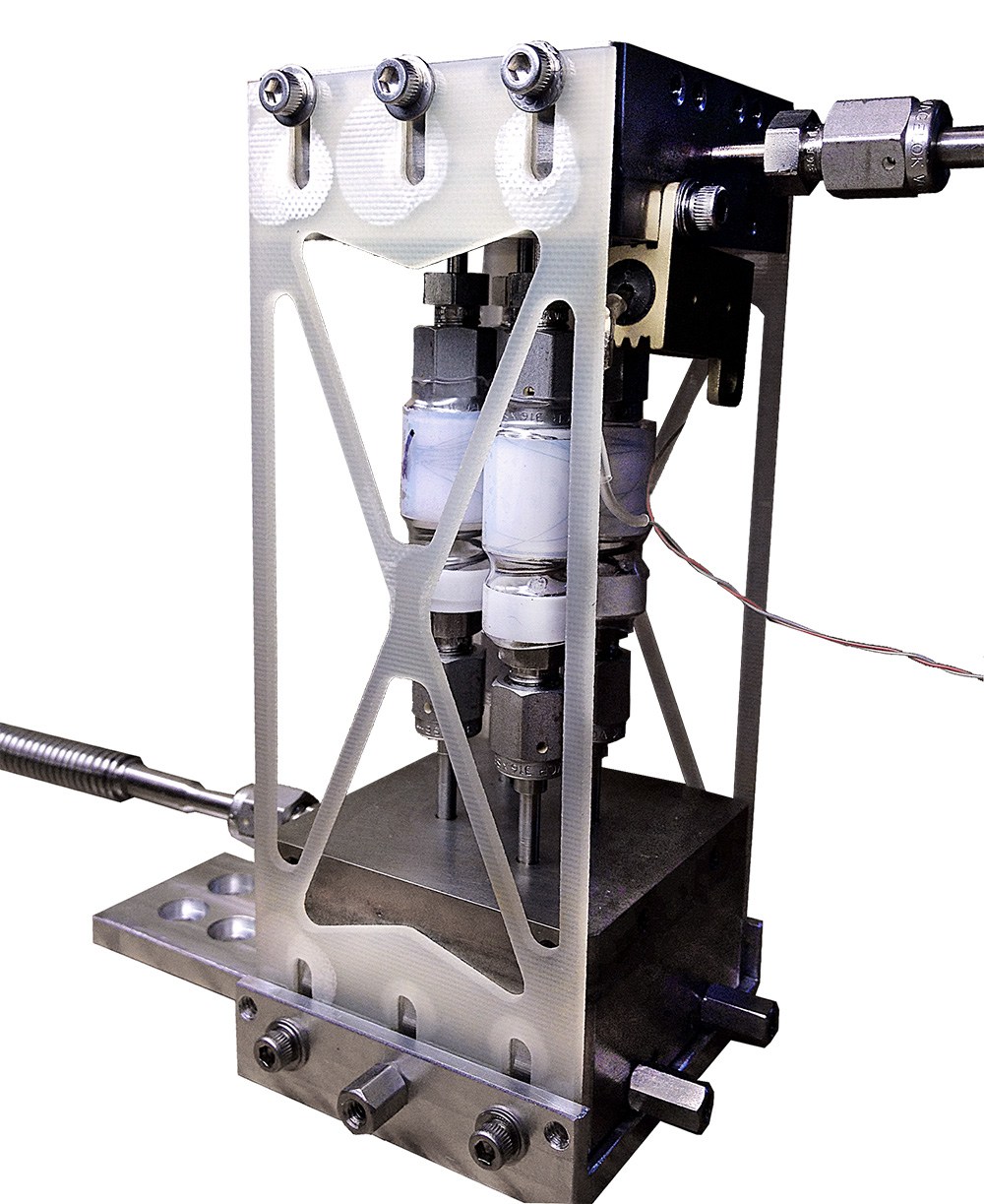}
\caption{A section view of the capillary assembly (left) and completed structure
  (right). Four capillaries connect the 4\,K main tank box (bottom) to the
  1.6\,K superfluid box (top). The double volume structure is supported by two
  \sfrac{1}{32}'' thick G10 flexures. The thermal load conducted through these
  flexures is negligible compared to the cooling power from the superfluid
  $^4\mathrm{He}$. Porous stainless steel Mott filters located below each
  capillary prevent ice and other dirt from entering and clogging this high
  impedance tubing. Superfluid helium exits the capillaries in the smaller of
  the two boxes (top) which is connected to the superfluid tank through 6'' long
  bellows tubing with a \sfrac{1}{8}'' diameter. \label{fig:cap}}
\end{figure}

The \spider capillary system (Figure \ref{fig:cap}), based on a classic
design\cite{Delong1970}, provides continuous flow of liquid $^4$He from the main
tank to the superfluid tank. The system is critical for a successful flight
because the hold time of the SFT, when fully charged, is only 4~days, compared
to the goal 20~day flight duration. The capillary system has been characterized
extensively during cryogenic testing of the flight cryostat. The system provides
approximately 100\,mW of cooling power to the superfluid tank, which is
sufficient to cool six telescopes down to 1.6\,K while sustaining daily fridge
cycles.  The net cooling power can be changed by simply altering the length of
the capillaries in a way that does not require any other change to the design.
This has allowed us to arrive at an optimal cooling power by having a collection
of interchangeable capsules. We find that measurements of 300\,K flow impedance
is an adequate predictor for superfluid flow rates.\cite{jeg_cap}

\subsection{Sub-Kelvin Systems}

\begin{figure}
\centering
\includegraphics[height=0.25\textheight]{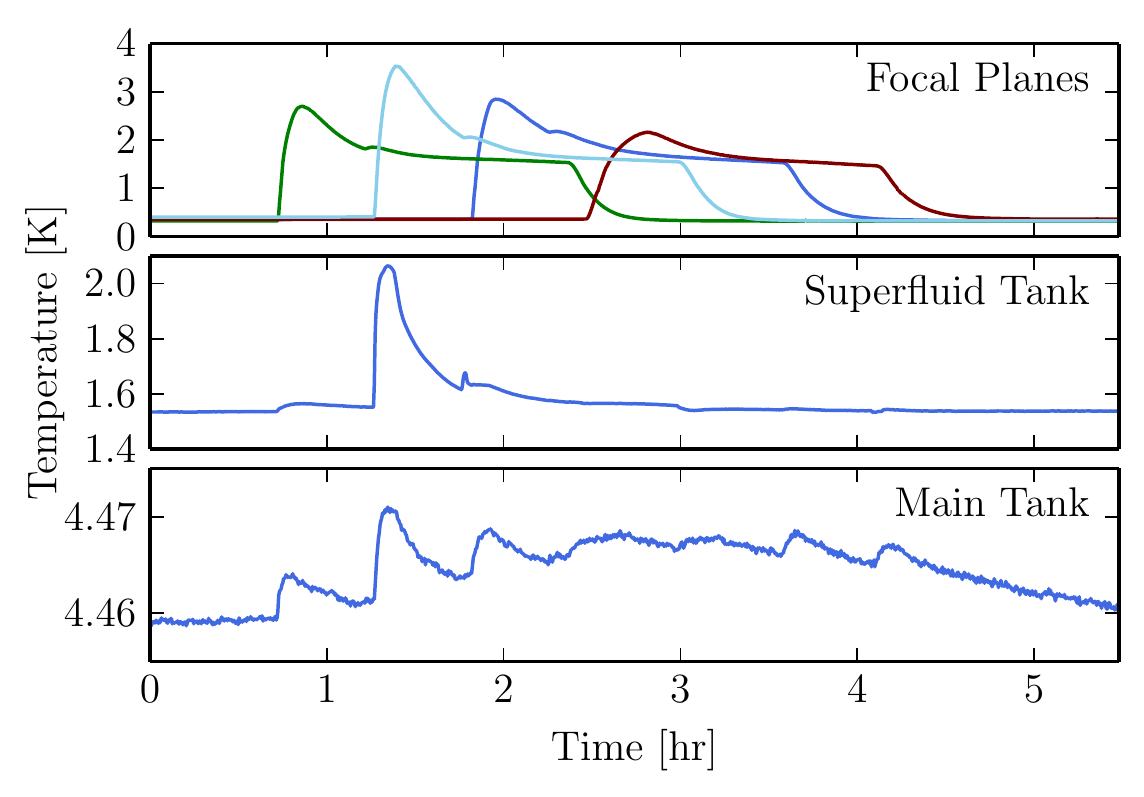}
\caption{A time stream of thermometry data while cycling four $^3$He
  refrigerators. The cycles are staggered half an hour apart, as seen at top, to
  maintain a manageable load on the SFT. The load on the MT is negligible, and
  though the SFT temperature increases, it returns to equilibrium quickly.  The
  large 0.5\,K rise in the SFT temperature at hour 1.3 is due to the proximity
  of the thermistor to that refrigerator's condensation point, which heats
  temporarily when the $^3$He gas is first released at the beginning of the
  cycle.}
\label{fig:fpus_cycles}
\end{figure}

Each receiver is cooled to 300\,mK with an independently controlled single-stage
closed-cycle $^3$He refrigerator made by Chase Cryogenics.  One such
refrigerator requires about two~hours to recycle at an elevation of 40$^\circ$,
and the system can comfortably handle cycling six such refrigerators with a wait
time of about 30\,min between each cycle start (Figure \ref{fig:fpus_cycles}).
The total cycling procedure takes about 4-5~hours to complete, but the focal
plane temperatures are not affected by neighboring cycles.  The hold times of
the fridges are typically in excess of 48~hours.  Accounting for non-observing
time for tuning and calibrations, we can make a pessimistic estimate of a duty
cycle of 85\%, which is in line with our design goal.

\subsection{Cooldown Procedure}

In addition to improvements to the radiative filtering, we have optimized the
close-up and cooling procedure to ensure optimal performance of the cryogenic
systems.  The initial evacuation of the vacuum chamber is critical for removing
water vapor and other particulates that can degrade the performance of the MLI
and introduce excess loading when the system is cold.\cite{bapat1990}  Upon
sealing up the vacuum vessel, we perform at least three purge cycles with dry
nitrogen gas; this has been shown to improve the ultimate vacuum pressure that
is reached prior to filling with cryogens.  Reaching vacuum pressures of
$\mathcal{O}(10^{-3})$\,torr prior to filling cryogens takes at least three days
due to the large volume and out-gassing surface area of the system.  Once a low
vacuum pressure is reached, the main tank is filled with $300-400$\,L of liquid
nitrogen to pre-cool the VCS layers and receivers; the time to equilibrium is
approximately another three days.  The SFT is maintained at a pressure of 5\,psi
above atmosphere using dry helium gas, to prevent the capillary lines from
clogging with LN$_2$ or water vapor, and cools radiatively during this period.
Finally, the LN$_2$ is removed by pressurizing the main tank, and replaced with
liquid helium.  The SFT is evacuated with a roughing pump and rapidly cools to
below the 2.2\,K $\lambda$-point of $^4$He once steady liquid flow is
established.  Final equilibrium temperatures are reached in another $3-4$ days,
at which point the refrigerators are ready to cycle for operation of the
detectors. In total, the vacuum pumping and cooldown to operational temperatures
requires about two weeks to complete.

The system will be launched with the detectors cold and operational, and with
the SFT evacuated.  A low-power diaphragm pump will be powered up shortly before
launch, to replace the roughing pump during ascent and maintain the SFT pressure
below $\sim$100\,torr.  Upon reaching float altitude ($\sim$35\,km), the
diaphragm pump will be bypassed in favor of the low-pressure atmosphere, by
commanding a pair of motorized bellows valves.

The cryogenic systems have been thoroughly tested in preparation for flight, and
found to perform adequately for our science goals.

\section{Telescope Performance}\label{sec:perf}

\subsection{Detector Characterization}\label{sec:det_perf}

The noise performance of the detectors has been well characterized in the flight
cryostat under flight-like loading conditions.  We expect the sky loading in
flight to be an effective 4-5\,K source (including the CMB and the atmosphere),
which can be simulated on the ground by covering the focal plane at the 4\,K
snout with an aluminum plate.  In a smaller test cryostat, flight-like loading
can be simulated using a temperature-controlled liquid $^4$He cold load attached
at the window aperture, and has been used to accurately measure instrument
loading and optical efficiency on the titanium
devices.\cite{art_thesis,rst_thesis} As an alternative to the cold load, a
removable curved mirror at the window can be used to present an effective
10-20\,K load, just under the saturation temperature of the titanium devices.

\begin{figure}
\centering
\includegraphics[width=0.5\textwidth]{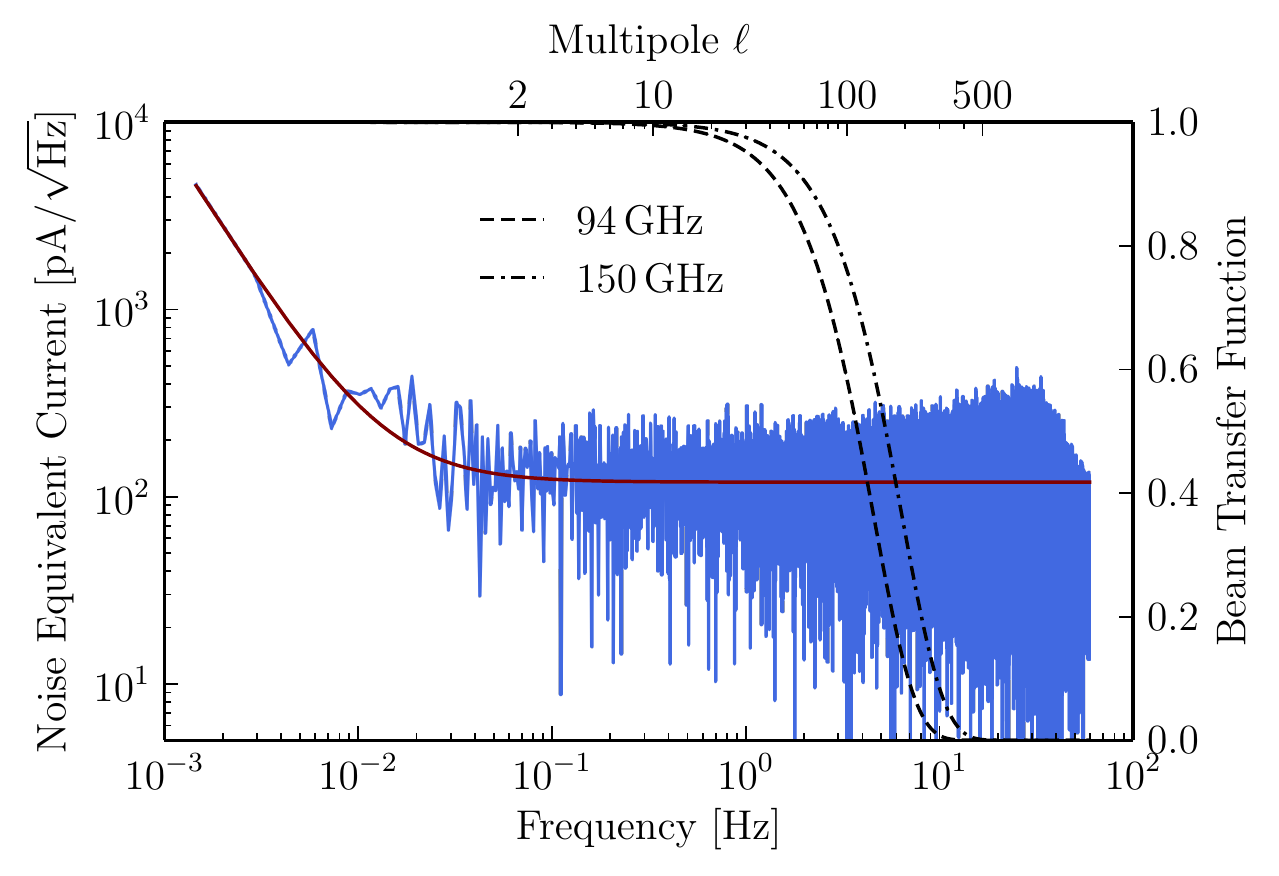}
\caption{Amplitude spectral density of a typical detector in current units, with
  a fit to a spectral shape including $1/f$ and Gaussian noise components.  For
  typical scan speeds of 6\,deg/s, the noise is relatively flat over the angular
  scales of interest ($10 < \ell < 300$), where $\ell \sim
  f\,v_\mathrm{scan}/180^\circ$.  Note that the excess amplitude above 10\,Hz is
  outside of the beam transfer function at these scan speeds.  This excess noise
  is believed to be due to the distributed heat capacity in the long meandered
  legs leading up to the bolometer island.\cite{obrient2014}}\label{fig:noise}
\end{figure}

The shape of the noise spectrum on a typical time-domain multiplexed device has
been measured in various test systems.\cite{art_thesis,rst_thesis}
The detector sample rate is limited by the capacitance of over 3\,meters of
cryogenic cabling between the multi-channel electronics (MCE) crate and the
focal plane.  The \spider switching rate between channels is 470\,kHz, which
gives a per-channel sample rate (after on-board decimation) of 120\,Hz.  A
typical noise spectrum is shown in Figure \ref{fig:noise}, acquired in the
flight system with the detector biased on the titanium transition at
$R\sim0.7R_n$.  This bias position is chosen to minimize noise-equivalent power
(NEP) on transition, as estimated from the noise plateau in the $\sim$2-8\,Hz
range.  At typical scan rates of 6\,deg/s the noise is relatively white over
the angular scales of interest ($10<\ell<300$).

\begin{figure}[b]
\centering
\includegraphics[height=0.23\textheight]{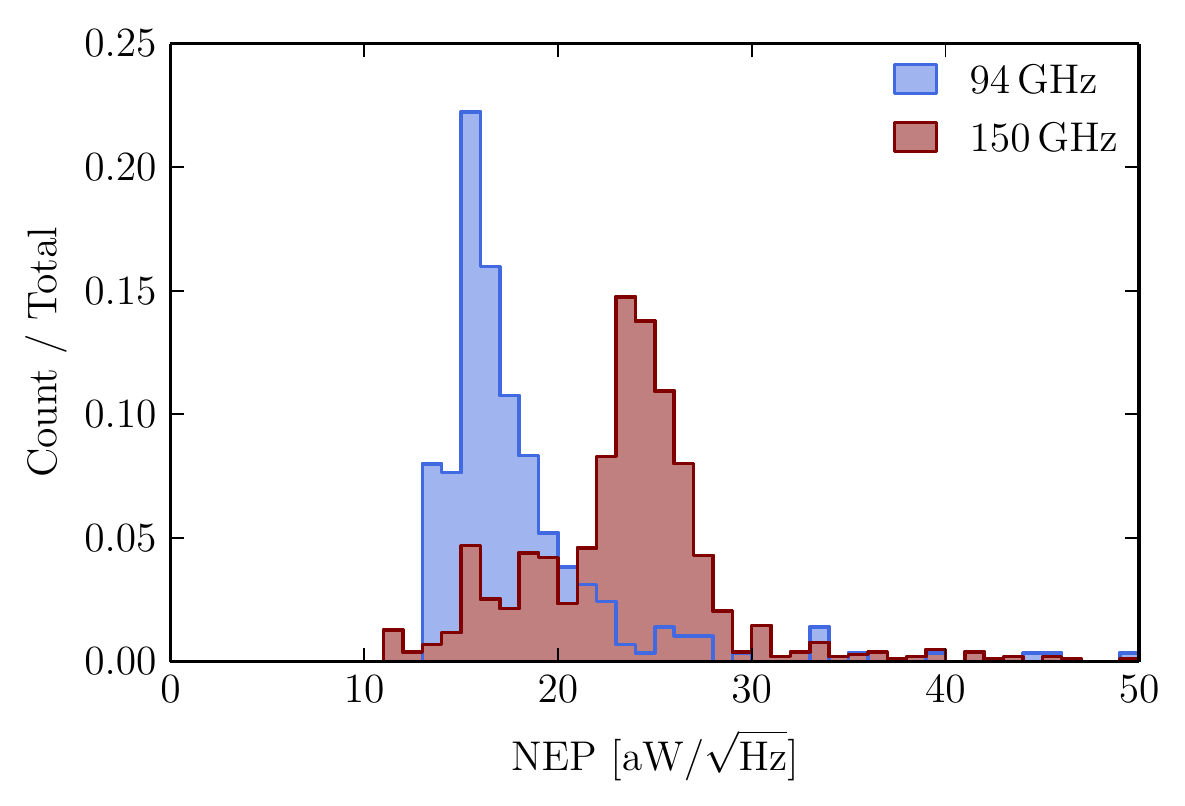}\qquad
\includegraphics[height=0.23\textheight]{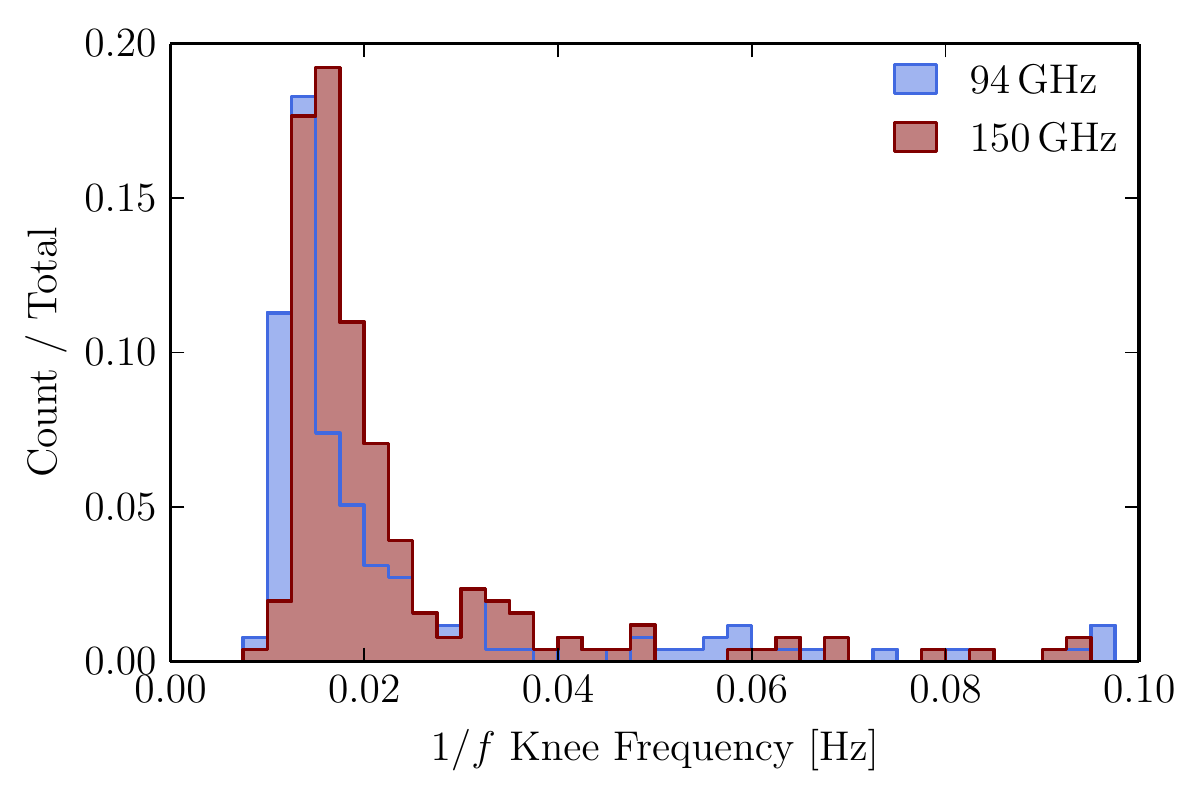}
\caption{({\itshape left}) Histogram of NEPs for a representative subset of
  94\,GHz and 150\,GHz detectors, measured during a period of 15 minutes of
  quiescent data acquisition while looking at a 4\,K load in the flight
  system. These NEPs correspond to NETs of about 120\,\ukrts and 140\,\ukrts,
  respectively. ({\itshape right}) Histogram of $1/f$ knee frequencies during
  the same acquisition period.}\label{fig:det_hist}
\end{figure}

To determine the signal-to-noise on the sky, the noise-equivalent current (NEI)
on the detector must be converted to noise-equivalent temperature (NET)
referenced to the input signal:
\begin{equation}
\mathrm{NET} = \mathrm{NEP} \, \frac{dT}{dP_\mathrm{opt}} = \mathrm{NEI} \, \frac{dP_\mathrm{opt}}{dI_s} \frac{dT}{dP_{opt}},
\end{equation}
where $I_s$ is the current signal, $dI_s/dP_\mathrm{opt}$ is the detector
responsivity to an optical signal, and $dP_{opt}/dT$ is the system optical
efficiency.  For fast detectors (loop gain $\mathcal{L}_I \gg 1$, time constant
$\tau \sim 1$\,ms), we can use the DC approximation in the large loop gain
limit, and the optical responsivity scales linearly with the electrical
responsivity\cite{irwin_hilton}:
\begin{equation}
\left.\frac{dI_s}{dP_\mathrm{opt}}\right\vert_{\omega=0} = \frac{1}{I_b R_L} \left( 2 \left.\frac{dI_s}{dI_b}\right\vert_{\omega=0} - 1 \right) = \frac{2 \eta\vert_{\omega=0} - 1}{I_b R_L},
\label{eq:resp}
\end{equation}
where $I_b$ is the bias current (typically $\sim$100\,$\mu$A), and $R_L$ is the
load resistor in the voltage-biased TES circuit (3\,m$\Omega$). This linear
relation has been verified for the \spider detectors with a chopped thermal
source placed at the aperture.  The second equality above defines the electrical
responsivity $\eta$, which can also be written in this limit in terms of the TES
resistance $R$ as
\begin{equation}
\eta\vert_{\omega=0} = -\frac{R_L}{R-R_L}.
\label{eq:eta}
\end{equation}
The electrical responsivity can be quickly measured by stepping the TES bias
current by a known small amplitude at a relatively low frequency ($\sim$1\,Hz),
and measuring the amplitude of the signal current.  These electrical ``bias
steps'' enable efficient monitoring of the detector NEP in flight, and are
discussed further in Section \ref{sec:det_mon}.  Histograms of typical plateau
NEP values and the knee frequency of the $1/f$ component are shown in Figure
\ref{fig:det_hist}, and indicate relatively quiet and stable performance of the
detectors.

\subsection{Optical Characterization}

\subsubsection{Beams}

\begin{figure}
\centering
\includegraphics[height=0.25\textheight]{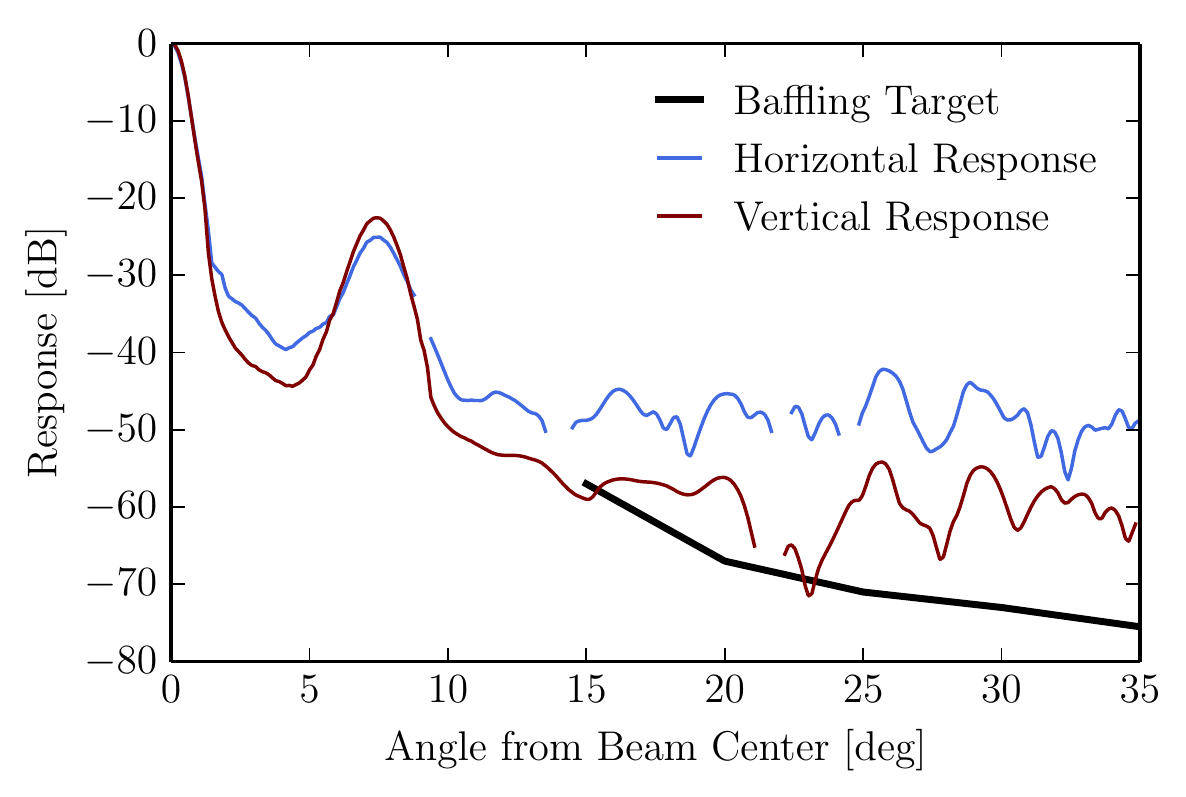} \qquad
\includegraphics[height=0.25\textheight]{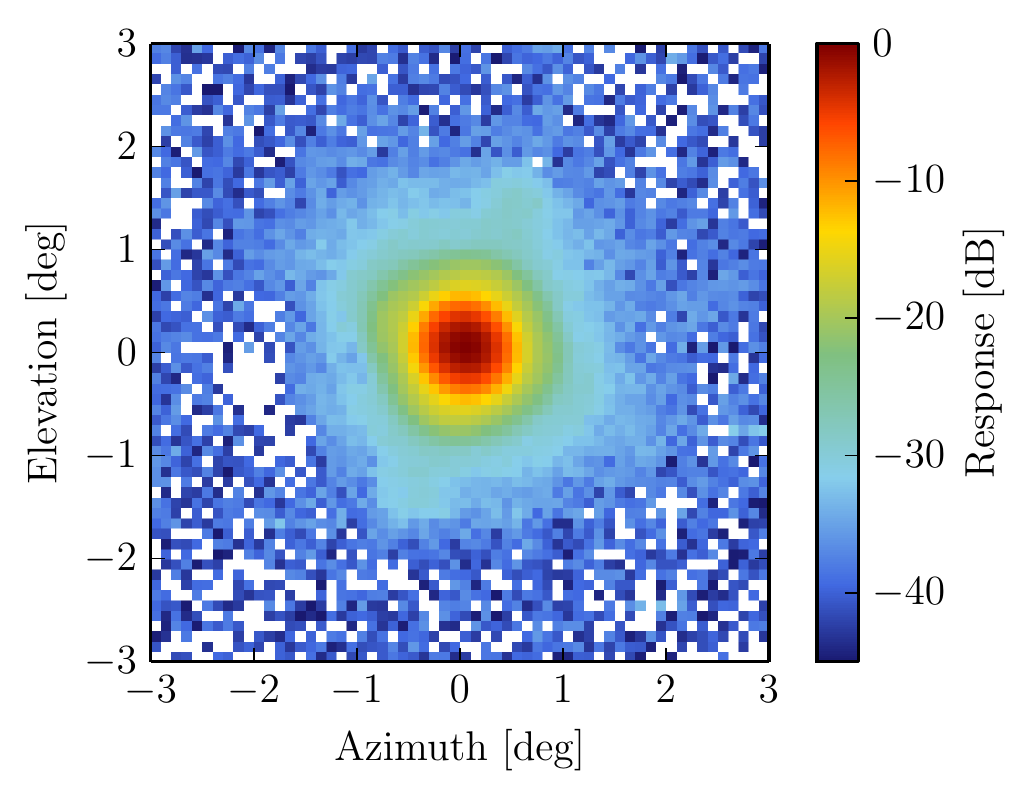}
\caption{(\textit{left}) A slice through a single 150\,GHz beam with both
  horizontal and vertical polarizations, acquired using a polarized source in
  the far field.  A ghost beam is evident 8$^\circ$ away from the beam center,
  due to reflections off of elements along the optical path.  The forebaffle
  attenuates the signal beyond about 10$^\circ$, so the sidelobes appear to be
  below the noise floor in this dataset.  The target level of sidelobe
  attenuation (black) was determined from a set of simulations of a \spider scan
  across a sky with Galactic emission\cite{fraisse2011_all}.  (\textit{right})
  Stacked beams on a 150\,GHz focal plane, with cross-talk beams between
  neighboring pixels visible above and below the main beam.  The orientation of
  these secondary beams indicate that the focal plane was rotated with respect
  to the elevation axis by 22.5$^\circ$.  These data were acquired during
  scanning tests of the fully integrated system in Palestine, TX, with the
  pointing solution reconstructed by integrating the 3-axis gyroscope
  signal.}\label{fig:beams}
\end{figure}

The detector main beams have been extensively measured in a test system, mounted
on an azimuth-elevation pointing system and scanning a thermal source near the
far field ($\sim$30\,m).  The beam widths have been measured to be 42\,arcmin at
94\,GHz and 30\,arcmin at 150\,GHz, with differential beam parameters (width,
ellipticity, centroid offset) at the 1\% level or below.\cite{rst_thesis}.

Sidelobes have been characterized using wide-angle scans pointed at an amplified
polarized source, and are currently limited by the brightness of the source.  A
slice through a single detector's beam (Figure \ref{fig:beams}) shows the
response of each polarization in a single physical pixel.  Also evident is a
``ghost'' beam about 8$^\circ$ from the beam center, which is a typical
consequence of reflections off of optical elements in a refractive
configuration.  This ghost has been measured at an amplitude of $\sim$1\%.  The
ghost location depends on the location of the pixel within the focal plane; this
particular ghost is for a pixel located about 3-4 degrees from the bore-sight.

Signal pick-up (``cross-talk'') between neighboring detectors has also been
characterized.  Figure \ref{fig:beams} shows a 150\,GHz beam produced by
stacking all detectors on the focal plane.  The upper and lower cross-talk beams
are visible at about the $-25$\,dB level, aligned with the focal plane axes
($-22.5^\circ$ relative to the elevation axis).

All of these beam effects have been
simulated\cite{mactavish2008,odea2010_all,fraisse2011_all} and are expected to
contribute negligible levels of spurious $B$-mode power at these amplitudes
(relative to an $r=0.03$ primordial signal).  Moreover, rotation of the HWP 
provides a robust way of minimizing sensitivity to these beam effects.  
Rotation of the polarization angle on the sky without displacing the beam 
centroid enables the creation of daily maps in which these beam effects 
contribute with opposite sign, and can thus be easily separated from the signal.

\subsubsection{Sidelobes and Baffling}\label{sec:baffle}

\begin{figure}
\centering
\includegraphics[width=0.5\textwidth]{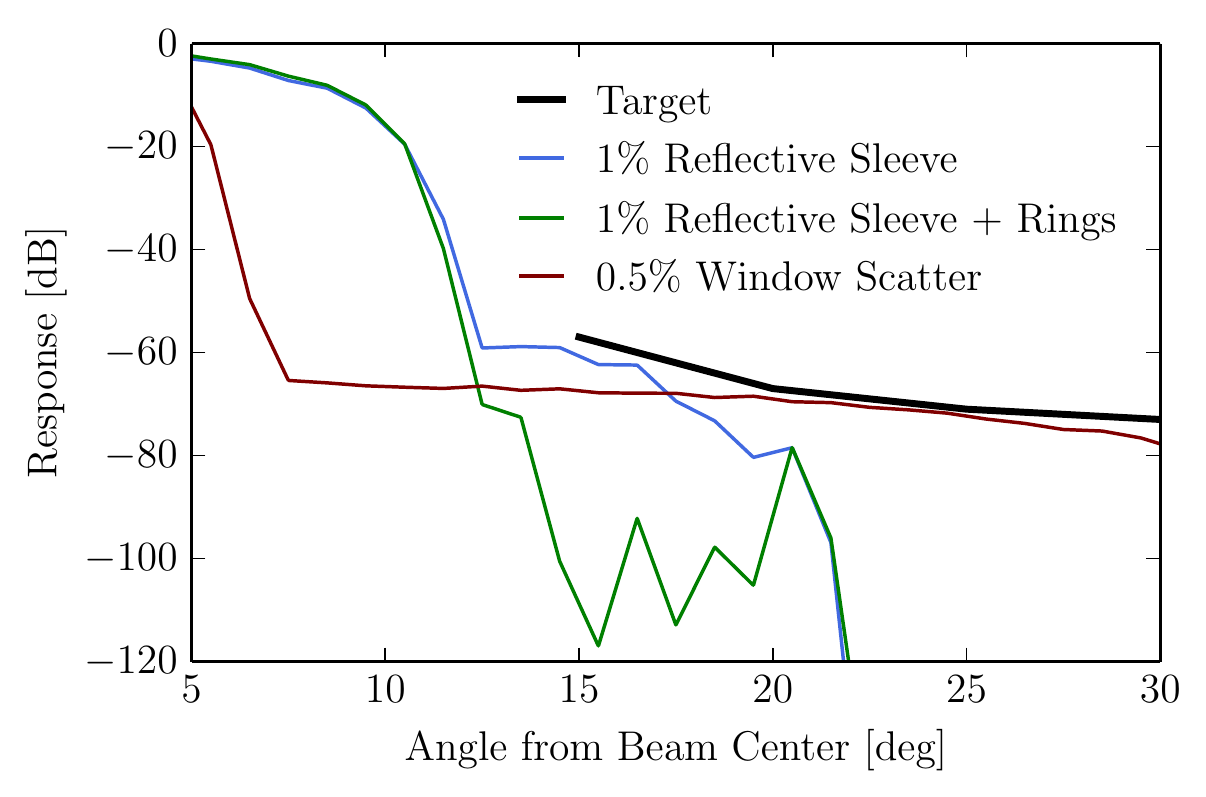}
\caption{Zemax\circledR\ modeling of attenuation of 100\% polarized sidelobes by
  internal and external baffling, at a pixel near the edge of the focal plane,
  where pickup is expected to be strongest.  The target level of attenuation
  (black) is also shown in Figure \ref{fig:beams}. If both horizontal and
  vertical polarizations are attenuated to at least this level, then excess
  sidelobe pickup is expected to be sub-dominant to a primordial $B$-mode signal
  at $r=0.03$.\cite{fraisse2011_all} Optical simulations of a 1\% reflective
  sleeve with (green) and without (blue) baffle rings along the length of the
  sleeve indicate adequate attenuation.  Simulations of 0.5\% Lambertian
  scattering off the window, include the forebaffle design shown in
  Figure~\ref{fig:insert} and a completely absorptive optics sleeve, indicate
  marginal performance beyond $\sim$20$^\circ$.  The combination of the external
  forebaffle and the blackened internal sleeve should limit the observed
  sidelobe pickup to acceptable levels, especially in the more realistic
  scenario that the sidelobes are only partially polarized.}\label{fig:baffle}
\end{figure}

Simulations of sidelobe pick-up of Galactic emission indicate that the
attenuation requirement for \spider is quite strict in order to reduce the
spurious $B$-mode contribution to well below a $r=0.03$ primordial
signal.\cite{fraisse2011_all}.  The attentuation target is shown in Figure
\ref{fig:baffle}.  To reach this target, we have built on the baffling scheme
typically used by the ground-based BICEP-style experiments.

The \spider forebaffles are designed to prevent contamination from polarized
sources, including reflections from the balloon and ground and wide-angle
Galactic emission.  The strongest constraint on the design of the forebaffles
comes from the spacing of the individual receivers.  To stay outside of the beam
and not interfere with neighboring baffles, simple conical forebaffles can
extend only 26 inches from the vacuum window.  Simulations show that this is
enough to reduce balloon and ground pick-up to acceptable levels, so a more
complicated design is not necessary.  The forebaffles are reflective, because
absorbing baffles were measured to contribute unacceptable levels of detector
loading (nearly 1\,pW) prior to the introduction of the 2\,K sleeve internal
baffles described below.  Lab measurements indicate less than 0.3\,pW of loading
from each forebaffle at both 94 and 150\,GHz.  The sidelobe power is dominated
by a combination of window scattering and optics sleeve reflections.  The
AR-coated UHMWPE windows scatter less than 0.5\% in our bands.  To reduce the
amount of scattered light reaching wide angles on the sky, the lower section of
each forebaffle features a zig-zagged portion, optimized to reflect some of
these rays back into the cryostat.

Reflections from the 2\,K sleeve have also been observed to contribute strongly
to the sidelobe power.  The optics sleeve has typically been blackened with
steel- and carbon-loaded STYCAST\circledR, but this coating has been found to be
as much as 80\% reflective for $\sim$2\,mm thicknesses at shallow angles.  A
surface at least 5\,mm thick would be required to reduce reflections to below
15\%.  To stay within mass constraints, we have instead covered the sleeve with
Eccosorb\circledR\ HR-10 material, which should ensure 1\% reflectivity at
microwave frequencies.  In addition to the strongly absorptive material, a
series of baffle rings have been placed in strategic locations along the length
of the optics sleeve, and optimized to reflect light back into the cryostat.
These baffle rings alone have been shown to reduce the loading on each receiver
by about 0.2\,pW.  The unique design of our forebaffle combined with these
internal baffle rings allows us to limit our sidelobes to the required levels,
as shown in Figure \ref{fig:baffle}.

\subsubsection{Bandpass and Optical Efficiency}

\begin{figure}
\centering
\includegraphics[width=0.5\textwidth]{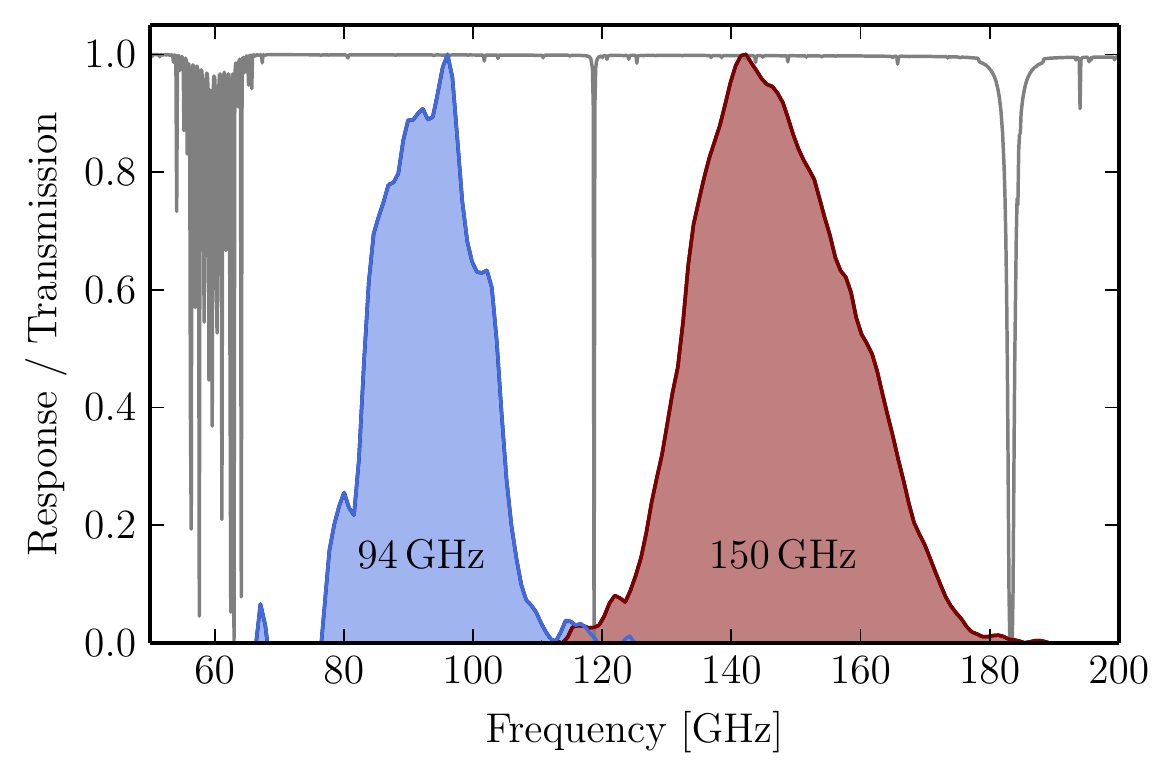}
\caption{Average spectral response of the 94\,GHz (blue) and 150\,GHz (red)
  detectors.  These spectra correspond to a 24\% bandwidth at both frequencies.
  The bands were placed to avoid the 60\,GHz and 180\,GHz water vapor resonances,
  and the 120\,GHz oxygen resonance that are present in the atmospheric
  transmission spectrum at 30\,km altitude (gray).}\label{fig:fts}
\end{figure}

The spectral bandpass of each detector is defined by both a low-pass filter
mounted at the aperture of the 2\,K magnetic shield, and a 3-pole LC filter
along the microstrip line leading to each bolometer.  The bandpass has been
measured using a compact polarized Fourier transform spectrometer (FTS) with a
chopped thermal source.  The average bandpass for both frequencies is shown in
Figure \ref{fig:fts}.  The bandwidth is about 24\% for both frequencies.  No
significant mismatch between polarizations has been observed.

The end-to-end optical efficiency of the detectors is determined by biasing on
the aluminum transition, measuring the change in bias power between a 300\,K and
a 77\,K (LN$_2$) load, and comparing to the expected $dP/dT$ for a blackbody
source with the observed bandpass.  Observed efficiencies are typically about
35-40\%, with some channels as high as 50\%.

\subsubsection{Polarimetric Calibration}

Cross-polar response has been limited to 0.5\% on typical detector pairs, whose
dominant source is multiplexer cross-talk\cite{bicep2_inst}.  This measurement
was done using a rotating polarized source (RPS) in the near field, peaked on
one detector at a time.  A complete revolution of the HWP at various fixed RPS
angles produced modulation of both detector polarizations, from which the
relative orientations of the detectors were determined.

Final polarization calibrations will be tested before deployment, and completed
just before launch.  We plan to use a rotating polarized grid at the window
aperture, to determine the orientations of all of the detectors relative to each
other, to the HWP, and to the cryostat reference frame.  The requirement for the
accuracy of this measurement is about 1$^\circ$ for both relative and absolute
angle calibrations,\cite{fraisse2011_all} which is easily achievable with the
HWP encoder system (accurate to 0.1$^\circ$).\cite{sab_thesis}

\section{Instrument Integration}\label{sec:integ}

Full integration of the \spider flight system was successfully completed
during the summer of 2013 at NASA's Columbia Scientific Ballooning Facility
(CSBF) in Palestine, TX.  This included assembly of the solar power systems,
integration with the NASA telemetry equipment, and integration of the detector
control systems with the flight commanding and downlink infrastructure.

\subsection{Power Systems}

The power systems for \spider are based largely on designs used by the BLAST
and BLASTpol instruments.\cite{blast_inst}  The electronics are divided into
an inner frame and an outer frame, which are electrically isolated from each
other.  The inner frame includes the detector readout electronics (one MCE
crate per receiver), the cryogenic housekeeping system, and the HWP control
system.  The outer frame includes all flight computers (one per receiver, and
two redundant control computers), peripherals (data storage pressure vessels,
serial hub, etc), the pointing control systems, and the line-of-sight
telemetry equipment.  The cryostat vacuum shell is the electrical ground for
the MCEs, and is carefully isolated from the outer frame electronics by means
of isolating DC-DC converters.

Two $4\times3$ arrays of solar panels (SunPower\circledR\ A-300) provide
independent power to the inner and outer frame systems.  Solar power from each
array is fed to a pair of series 12\,V lead-acid batteries (Odyssey PC1200), via
a solar charge controller (MorningStar\circledR\ TriStar MPPT-60).  This power
system was fully assembled and tested in summer 2013, and is expected to provide
up to 2\,kW of solar power during flight, with an efficiency of up to 21\%.

\subsection{Payload Control and Telemetry}

On-board control of the \spider payload is handled by a pair of redundant
flight computers, which communicate with each of the receiver computers, the
pointing motors and sensors, and the housekeeping electronics.  Data are
stored on redundant systems, with at least three copies distributed among
various models of spinning hard disks (enclosed in pressure vessels) and solid
state drives, to minimize the probability of data loss due to drive failure.
A 25\,MHz clock synchronizes data between the flight computers and the six
MCEs.

Communication with the payload from the ground is facilitated by a Support
Instrumentation Package (SIP), provided by CSBF.\cite{csbf_sip}  The SIP
allows access to the TDRSS and Iridium satellite networks for communication,
as well as a line-of-sight (LOS) transmitter for use during the first 24-36
hours of the flight.  Data transmission rates are up to 1\,Mbps via LOS, and
up to 6\,kbps over the horizon, with at the very least a 255-byte packet 
transmitted every $\sim$15 minutes.  These data rates are too low to allow
transmission of the full \spider dataset ($\sim$12\,Mbps), so only a subset of
the data are transmitted during the flight for monitoring.  The SIP was
installed and tested during the 2013 integration campaign.

\subsection{Detector Tuning and Monitoring}

\spider will be the first payload to fly a time-domain multiplexed detector 
readout system.  In order to monitor and control the over 2000 individual
channels given limited downlink bandwidth, we have created a custom
infrastructure for quickly determining the health of each receiver in flight.
This system includes tuning of the SQUID readout electronics, tuning of the
detector responsivity, and monitoring of the detector stability throughout the
flight.

The monitoring software runs independently on each MCE computer, and handles
acquisition of data onto each of three redundant data drives, communication 
with the flight computer, and tuning and monitoring of the readout system.  The
program stores a copy of all runtime parameters on each data drive, from which
it can recover the last known working state in the event of loss of
communication or power failure.

\subsubsection{SQUID Tuning}

The detector resistances are read out by a 3-stage system of SQUIDs that
inductively couple to the TES bias circuit.\cite{battistelli2008}  The 
detectors are arranged into 33 rows and 16 columns in the multiplexer circuit. 
Each column of SQUIDs and TESs is biased simultaneously, while each pixel is 
sampled by the multiplexer at 14.3\,kHz.

The voltage across each SQUID is a periodic function of the magnetic flux 
through the SQUID loop, with period equal to the magnetic flux quantum.
To tune the SQUIDs, the feedback and bias current at each SQUID stage is swept 
to maximize both linearity and gain of the response over the dynamic range.  
Once each SQUID stage has been properly tuned by row (first stage) or by 
column (second and last stages), a lock point is further optimized for each 
individual channel, based on the composite response through all three SQUID 
stages.\cite{battistelli2008_tuning}

For ground-based operation, this tuning procedure can be vetted by plotting the
response curves for each SQUID and verifying by eye that the chosen lock points
are correctly placed along each curve.  To remove the user from this procedure,
the \spider tuning algorithm reduces the SQUID response curves and lock points
to a set of statistics (period, slope, amplitude, lock position) for each stage.
These statistics are compared against a reference tuning, assuming some allowed
thresholds in variation of each statistic.  The tuning is successful if the
number of failures is below a maximum threshold at each stage.  The tuning
algorithm will proceed to the next stage on success, or determine some
parameters to adjust and repeat on failure.  One typical failure mode is a shift
in the optimal SQUID bias due to drifts in the SQUID module temperature (common
when the liquid cryogen level is low); the algorithm determines the appropriate
bias sweep range and finds the optimal bias to maximize the SQUID gain.

Once the SQUIDs are tuned and locked, the detectors are operated in a closed
integral servo loop, with the SQUID feedback adjusted in response to changes in
the TES resistance.  The gain of the integral term has been determined based on
the performance of each column to quickly compensate the TES response without
inducing runaway oscillations.

A full tuning procedure, to determine both bias and feedback positions for all
SQUID stages, typically takes about five minutes per receiver, and will be
carried out roughly once per day.  To tune just the feedback positions takes
about one minute.  The latter procedure may be necessary more frequently if
magnetic field drifts or cosmic rays should cause the SQUIDs to become unlocked.
A new tuning can be triggered by command or if a threshold number of detectors
slip out of lock, discussed further in Section \ref{sec:det_mon}.

\subsubsection{TES Bias Tuning}\label{sec:det_tune}

\begin{figure}
\centering
\includegraphics[width=0.5\textwidth]{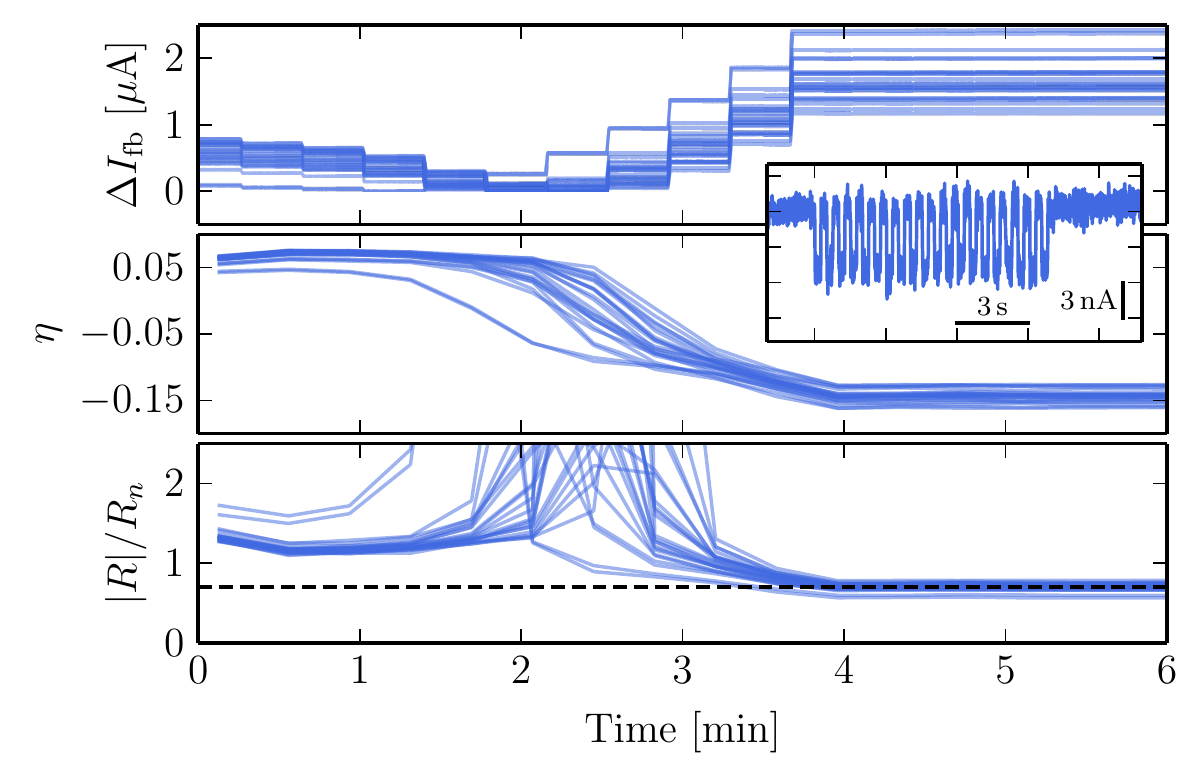}
\caption{Illustration of automatic detector biasing algorithm. ({\itshape top}) 
  Feedback signal for detectors in a single column as the TES bias is adjusted.
  ({\itshape inset}) At each bias level, the bias is pulsed by 1\,ADU at 2\,Hz 
  and the feedback response is recorded.  ({\itshape middle}) The measured 
  detector response $\eta$ is negative when the detectors are on transition. 
  ({\itshape bottom}) The effective resistance of each detector (using 
  equation~\ref{eq:eta}), approaches the column average target $R = 0.7 R_n$
  after several steps.}\label{fig:bias_steps}
\end{figure}

Tuning the TES bias can be done by acquiring a load curve for each channel,
which is a measure of the current response as a function of bias current.  In
order to bias the TES on transition starting from a superconducting state, the
focal plane must be thermally ``kicked'' onto transition, because the bias
circuit cannot provide enough current to surpass the critical current necessary
to drive the detectors normal (above the superconducting transition).  Once the
detectors on each column are biased and normal, the column bias is ramped down
through the superconducting transition, and a bias is chosen such that the NEP
on transition is minimized, as discussed in section \ref{sec:det_perf}.  The
optimal bias is typically when the detectors are high in the transition, where
excess amplifier noise is minimal, so we choose a bias such that the column
average TES resistance is 0.7 of the normal resistance $R_n$.

Acquisition of a full load curve is typically a disruptive procedure, and
hysteresis in the focal plane temperature after a kick may indicate a slightly
lower bias than is actually necessary.  To bias the detectors more efficiently
and reliably, we instead take local measurements of the load curve by commanding
small periodic ($\sim$1\,Hz) steps on the TES bias lines, and measure the
amplitude of the signal response (see Figure \ref{fig:bias_steps} inset).  The
amplitude of the bias step response on transition ($\eta$) is related to the TES
resistance by equation \ref{eq:eta} (Section \ref{sec:det_perf}).  Because
$\eta$ is a smooth function of the TES resistance, the TES bias can be tuned by
servoing off of the column averaged electrical responsivity.  Figure
\ref{fig:bias_steps} illustrates how this bias tuning procedure is implemented
on a single column.  This servoing procedure also naturally allows monitoring of
the TES responsivity throughout the flight, as discussed in the next section.

\subsubsection{Detector Monitoring}\label{sec:det_mon}

Once the SQUIDs and detectors are tuned and normal data acquisition has begun, a
monitoring program ensures that detector performance remains stable.  For each
receiver, a set of four statistics are calculated at the data frame rate:

\begin{quote}
\begin{description}
\item[Bias step response] The TES bias is stepped as discussed in
  Section~\ref{sec:det_tune}, and the responsivity is stored.  If the
  responsivity has changed according to a threshold on the column averaged TES
  resistance, then the TES bias is adjusted appropriately.  Bias steps are
  acquired at every scan turnaround, and adjusted if necessary roughly every
  hour.
\item[Mean] A channel in oscillation will show a DC mean level that has deviated
  significantly from zero, indicating a poorly locked channel.  This is also a
  good indicator of a large response to a cosmic ray event or other strong
  signals.
\item[Noise current] The noise current is estimated by taking the RMS of each
  detector time stream after applying an IIR filter to limit the bandwidth to
  the plateau region (typically chosen as 2-8\,Hz).  Combined with the bias step
  response, this gives an instantaneous estimate of the NEP per channel.
\item[RMS] The full-bandwidth noise over a period of about a minute is a good
  indicator of higher frequency noise, such as RF pickup, or low-frequency
  $1/f$ noise.
\end{description}
\end{quote}

\noindent  Because full detector time streams are largely unavailable after the
LOS period, these statistics provide a means of monitoring detector behavior
from the ground, and responding as necessary.  To minimize the bandwidth
required to transmit these statistics, they are reduced to 8-bit numbers and
multiplexed into a single downlink channel.  Along with these statistics, each
detector is assigned a flag, to which the software is able to respond
automatically:

\begin{quote}
\begin{description}
\item[Superconducting / Low / OK / High / Normal] The location on transition is
  determined from the bias step response.  The bias is adjusted periodically if
  too many channels have slipped off of transition.  If too many detectors have
  latched superconducting, then the detectors are kicked back onto transition as
  necessary.
\item[Clamped] In an attempt to prevent oscillatory behavior from affecting
  neighboring pixels via their common electrical ground, a ``clamping''
  procedure has been implemented at the firmware level on the MCE.  This locks
  the integral term in the error loop to effectively stop the servo before the
  oscillations can disturb the neighbors.  A clamped detector is indicated by a
  signal time stream that is locked at a constant, non-zero value.
\item[Oscillating] A detector with particularly large variation is flagged as 
  oscillating if it is not caught by the clamping procedure.  The servo loop can 
  be reset (the integral term re-zeroed) to attempt to stop the oscillatory 
  behavior; if this fails, the gain of the integral servo can be adjusted.
\item[Frail] A detector flagged as frail will be reset with a reduced integral 
  servo gain to attempt to stop oscillatory behavior.
\item[Dead] If oscillation continues despite the lower gain, the detector can 
  be turned off altogether (zero gain) and flagged as dead.
\end{description}
\end{quote}

Thresholds are set for the number of flagged channels; if these thresholds are
exceeded, then a new SQUID tuning may be requested, or the MCE may be reset
altogether to the last known working state.  When only low bandwidth channels
are available for downlink, then the statistics are further reduced to a column
average, or a simple count of detectors in each flagged state.

This novel detector tuning and monitoring scheme is now used continuously during
lab testing, and will enable efficient monitoring and response by the ground
crew during flight.

\section{Conclusions}\label{sec:concl}

The \spider instrument is fully integrated and ready to deploy to McMurdo
Station, Antarctica for a December 2014 launch.  The telescope performance has
been extensively characterized in the flight system.  We will observe a large
fraction of the southern Galactic sky in an effort to detect the inflationary
$B$-mode signal at degree angular scales.  The BICEP2 experiment has observed
$B$-mode power on the sky at these scales,\cite{bicep2_spect} and \spider's
spatial, angular and frequency coverage make it an ideal instrument for further
characterizing the signal's angular spectrum, isotropy and frequency spectrum
across the sky.

\begin{acknowledgments}
The \spider collaboration gratefully acknowledges the
support of NASA (award numbers NNX07AL64G and NNX12AE95G), the Lucille and David
Packard Foundation, the Gordon and Betty Moore Foundation, the Natural Sciences
and Engineering Research Council (NSERC), the Canadian Space Agency (CSA), and
the Canada Foundation for Innovation. We thank the JPL Research and Technology
Development Fund for advancing detector focal plane technology.  W.\,C.\ Jones
acknowledges the support of the Alfred P. Sloan Foundation.  A.\,S.\ Rahlin is
partially supported through NASAs NESSF Program (12-ASTRO12R-004).  J.\,D.\
Soler acknowledges the support of the European Research Council under the
European Union's Seventh Framework Programme FP7/2007-2013/ERC grant agreement
number 267934.

Logistical support for this project in Antarctica is provided by the U.S.
National Science Foundation through the U.S. Antarctic Program. We would also
like to thank the Columbia Scientific Balloon Facility (CSBF) staff for their
continued outstanding work.
\end{acknowledgments}

\bibliographystyle{spiebib}
\bibliography{spider_integration_spie2014}

\end{document}